\documentclass[12pt,preprint]{aastex}
\newcommand{\beq}{\begin{equation}}
\newcommand{\beqa}{\begin{eqnarray}}
\newcommand{\eeq}{\end{equation}}
\newcommand{\eeqa}{\end{eqnarray}}
\newcommand{\simg}{\gtrsim}
\newcommand{\siml}{\lesssim}
\newcommand{\meszaros}{M${\acute {\rm e}}$sz${\acute {\rm a}}$ros}
\shorttitle{Radio afterglows of high redshift GRBs and HNe}
\shortauthors{Ioka \& \meszaros}
\begin{document}
\title{
Radio Afterglows of Gamma-Ray Bursts and Hypernovae at High Redshift,
and their Potential for 21-cm Absorption Studies
}
\author{
Kunihito Ioka$^{1}$
and Peter \meszaros$^{1,2,3}$
}
\affil{$^{1}$Physics Department and Center for Gravitational Wave Physics, 
Pennsylvania State University, University Park, PA 16802}
\affil{$^{2}$Department of Astronomy and Astrophysics, 
Pennsylvania State University, University Park, PA 16802}
\affil{$^{3}$Institute for Advanced Study, Princeton, NJ 08540}

\begin{abstract}
We investigate the radio afterglows of gamma-ray bursts (GRBs)
and hypernovae (HNe) at high redshifts and quantify their 
detectability, as well as their potential usefulness for 21 cm 
absorption line studies of the intergalactic medium (IGM) and 
intervening structures.
We examine several sets of source and environment model 
parameters that are physically plausible at high redshifts.
The radio afterglows of GRBs would be detectable out to $z \sim 30$,
while the energetic HNe could be detectable out to $z \sim 20$
even by the current Very Large Array (VLA).
We find that the 21 cm absorption line due to the diffuse neutral 
IGM is difficult to detect
even by the proposed Square Kilometer Array (SKA),
except for highly energetic sources.
We also find that the 21 cm line due to 
collapsed gas clouds with high optical depth may be detected
on rare occasions.
\end{abstract}


\section{Introduction}

Gamma-Ray Bursts (GRBs) are potentially powerful probes of the 
early universe \citep[e.g.,][]{miralda98,meszaros03,barkana04}.
The substantial evidence that massive stellar collapses lead to
long duration GRBs suggests that these outbursts probably occur
up to the highest redshifts of the first generation of stars
\citep[c.f.,][]{meszaros02,zhang03}. Their high luminosities 
make them detectable in principle out to 
redshifts $z \sim 100$ \citep{lamb00},
while their afterglows may be observable up to
$z \sim 30$ in the infrared, X-ray \citep{ciardi00,gou04}
and radio bands \citep{ioka03,inoue04}.
About 25\% of all GRBs detected by the upcoming Swift
satellite are expected to be at $z>5$ \citep{bromm02a}.
Such high redshift GRBs may have already been detected by the
BATSE satellite, as suggested by the empirical distance
indicators \citep{fenimore00,norris00,murakami03,yonetoku04}
and their theoretical interpretation \citep[e.g.,][]{ioka01,yamazaki04}.
The first generation of stars are expected to be very massive
\citep{abel02,bromm02,omukai03}, so that their outcome might be 
GRBs which are an order of magnitude more luminous than their low
redshift counterparts \citep[e.g.,][]{fryer01,heger03}.

One of the major uncertainties in modern cosmology is the 
reionization history of the universe, which has significant 
implications for the formation of the first stars and their feedback
on the subsequent evolution \citep[e.g.,][]{barkana01,miralda03}.
The analysis of Ly$\alpha$ spectra in the highest redshift quasars 
indicate that the reionization has essentially reached completion
at around $z \sim 6$ \citep[e.g.,][]{fan02},
while the WMAP polarization data imply an onset of the reionization 
process at much higher redshifts $z \sim 17 \pm 5$
\citep{kogut03,spergel03}.
Although other methods, such as the temperature of the 
Ly$\alpha$ forest \citep{theuns02,hui03},
the $\rm H_{II}$ region size around quasars \citep{wyithe04,mesinger04},
and the luminosity function of the Ly$\alpha$ emitters
\citep{stern04,malhotra04} also constrain the reionization history,
it is still unclear even whether the reionization is a continuous 
process or whether it occurred in two episodes \citep{cen03,wyithe03}.

21 cm tomography would be one of the promising ways to reveal
the reionization history of the universe\footnote{
Other methods using GRBs may include
the infrared spectroscopy \citep{miralda98,barkana04}
or photometry \citep{inoue04b}
and the dispersion measure \citep{ioka03,inoue04}.} \citep{tozzi00,shaver99},
by mapping the neutral hydrogen in the spatial and redshift space
via the redshifted 21 cm line \citep{madau97,scott90,hogan79}.
The 21 cm line is associated with the transition between
the singlet and triplet hyperfine levels of the hydrogen atom 
at the ground state \citep[e.g.,][]{field59a}.
The 21 cm line may be detectable in emission or absorption
against the cosmic microwave background (CMB)
\citep[e.g.,][]{ciardi03,furlanetto04,zaldarriaga04,iliev02}.
Alternatively, the 21 cm line may be also observed in absorption
against luminous background sources \citep{carilli02,furlanetto02}.
But this method requires luminous sources of radio continuum
at high redshifts, which remain largely speculative. 
The GRB radio afterglows would be a natural candidate.
However their radio fluxes at high redshift have not been 
estimated in detail. \citet{furlanetto02} suggested that GRB 
afterglows are too dim, but they did not explicitly calculate 
the self-absorption frequencies, nor did they include the 
reverse shock emission. Another type of related energetic
radio source which has not been investigated in this respect are 
hypernovae (HNe), which may be as frequent as GRBs.

In this paper, we investigate the GRB and HN radio afterglows at 
high redshifts, including also the possibly of more energetic versions 
of these objects which may be associated with a population III of stars.
We first estimate the maximum redshifts out to which these afterglows could
be detected with the Very Large Array (VLA),\footnote{http://www.nrao.edu/}
Low Frequency Array (LOFAR),\footnote{http://www.lofar.org/}
and Square Kilometer Array (SKA).\footnote{http://www.skatelescope.org/}
This can be important for broadband afterglow fits aimed at determining 
total energies and physical parameters \citep[e.g.,][]{panaitescu02} of 
bursts, out to very high redshifts.
We also quantify the detectability of the 21 cm absorption line in the 
afterglows. Since the afterglow model parameters, such as the energy and
duration of the GRBs as well as the ambient density may be different at 
high redshifts, we examine several possibilities of the model parameters.
However, even with the SKA, which has the best sensitivity among the 
proposed telescopes, and with our improved spectral-temporal  evolution
model, we find that the chances are low for the detection of 21 cm 
absorption lines from GRBs and HNe, except in the case of exceptionally 
energetic sources. This quantification exercise should be useful for the 
assessment and planning of future observing projects with advanced 
facilities and detectors.

The paper is organized as follows.
In \S~\ref{sec:flux}, we calculate the GRB and HN radio afterglows 
at high redshift, based on the afterglow model summarized 
in \S~\ref{sec:model}.
By comparing the afterglow fluxes with the VLA, LOFAR and
SKA sensitivities, we discuss the maximum redshift out to which the
radio afterglow of these sources would be detectable.
We calculate the 21 cm line absorption in the IGM and intervening
structures at $z \simg 6$ in \S~\ref{sec:z>6} and 
at $z \siml 6$ in \S~\ref{sec:z<6}. Our conclusions are 
summarized in \S~\ref{sec:con}. 
Throughout the paper we use a $\Lambda$CDM cosmology
with $(\Omega_{m}, \Omega_{\Lambda}, \Omega_{b}, h)
=(0.27, 0.73, 0.044, 0.71)$ \citep{spergel03}.

\section{Radio fluxes of high redshift afterglows}\label{sec:flux}

We model the GRB afterglow source as a relativistic shell 
expanding into a homogeneous interstellar medium (ISM)
with particle number density $n$ at redshift $z$.
The shell initially has an isotropic equivalent energy $E$,
a Lorentz factor $\gamma_{0}$, an opening half-angle $\theta$
and a width in the source frame $c T (1+z)^{-1}$, 
where we assume $\gamma_{0}^{-1}<\theta$ and
the shell width being related to the observed GRB duration $T$
\citep{kobayashi97}.
The true energy is given by
$E_{j} = \theta^{2} E/2$.
Two shocks are formed: a forward shock heating the ISM
and a reverse shock decelerating the shell.
At these shocks electrons are accelerated and magnetic fields are amplified,
leading to the synchrotron afterglow emission.
We assume that accelerated electrons have a power-law distribution
of the Lorentz factor $\gamma_{e}$ as $N(\gamma_{e}) d\gamma_{e}
\propto \gamma_{e}^{-p} d\gamma_{e}$ for $\gamma_{e}>\gamma_{m}$
\citep{sari98b}.
We also assume that fractions $\epsilon_{e,f}$ $(\epsilon_{e,r})$ 
and $\epsilon_{B,f}$ $(\epsilon_{B,r})$ of the shock energy go
into electrons and magnetic fields, respectively, at the forward (reverse) shock,
where the subscripts $f$ and $r$ indicate the forward and reverse shock, 
respectively.
We also assume adiabatic shocks.

Under the above assumptions, 
we can calculate the spectra and light curves of the afterglows
as summarized in \S~\ref{sec:model}.
There are $10$ model parameters:
$E$, $\theta$, $n$, $p$, 
$\epsilon_{e,f}$, $\epsilon_{B,f}$,
$\epsilon_{e,r}$, $\epsilon_{B,r}$,
$\gamma_{0}$, $T/(1+z)$,
but we assume $\epsilon_{e,f}=\epsilon_{e,r}$ for simplicity, 
so the actual number of parameters is 9.
We consider the possible difference between 
the magnetic field in the forward and reverse shocks, 
by using ${\cal R}_{B}=(\epsilon_{B,r}/\epsilon_{B,f})^{1/2}$
in equation (\ref{eq:rB}) instead of $\epsilon_{B,r}$,
since the ejected shell may be endowed with magnetic fields
from the central source \citep[e.g.,][]{zhang03b}.
We also take the sideway expansion of the jet,
the non-relativistic regime, and the reverse shock emission
into account (see \S~\ref{sec:model}).

In the following we examine several sets of model parameters
that are physically motivated at high redshift,
as summarized in Table~\ref{tab:model}.
We estimate the maximum redshift out to which these afterglows
could be detected with the VLA, LOFAR and SKA.
We also evaluate the peak fluxes of the afterglows at 
the observed frequency around $\nu \sim 100$ MHz
to discuss the detectability of the 21 cm absorption line
later in \S~\ref{sec:z>6} and \S~\ref{sec:z<6}.

\subsection{Standard GRB case}\label{sec:standard}

First we adopt typical parameters inferred
from the broadband afterglow fitting
\citep[e.g.,][]{panaitescu02,zhang03b}:
the isotropic equivalent energy $E=10^{53}$ erg,
the opening half-angle $\theta=0.1$,
the ISM density $n=0.1$ cm$^{-3}$,
the spectral index $p=2.2$,
the initial Lorentz factor $\gamma_{0}=200$,
the duration in the source frame $T/(1+z)=10$ s,
and the plasma parameters $\epsilon_{e}=0.1$,
$\epsilon_{B}=0.01$, ${\cal R}_{B}=1$.

\begin{figure}
\plotone{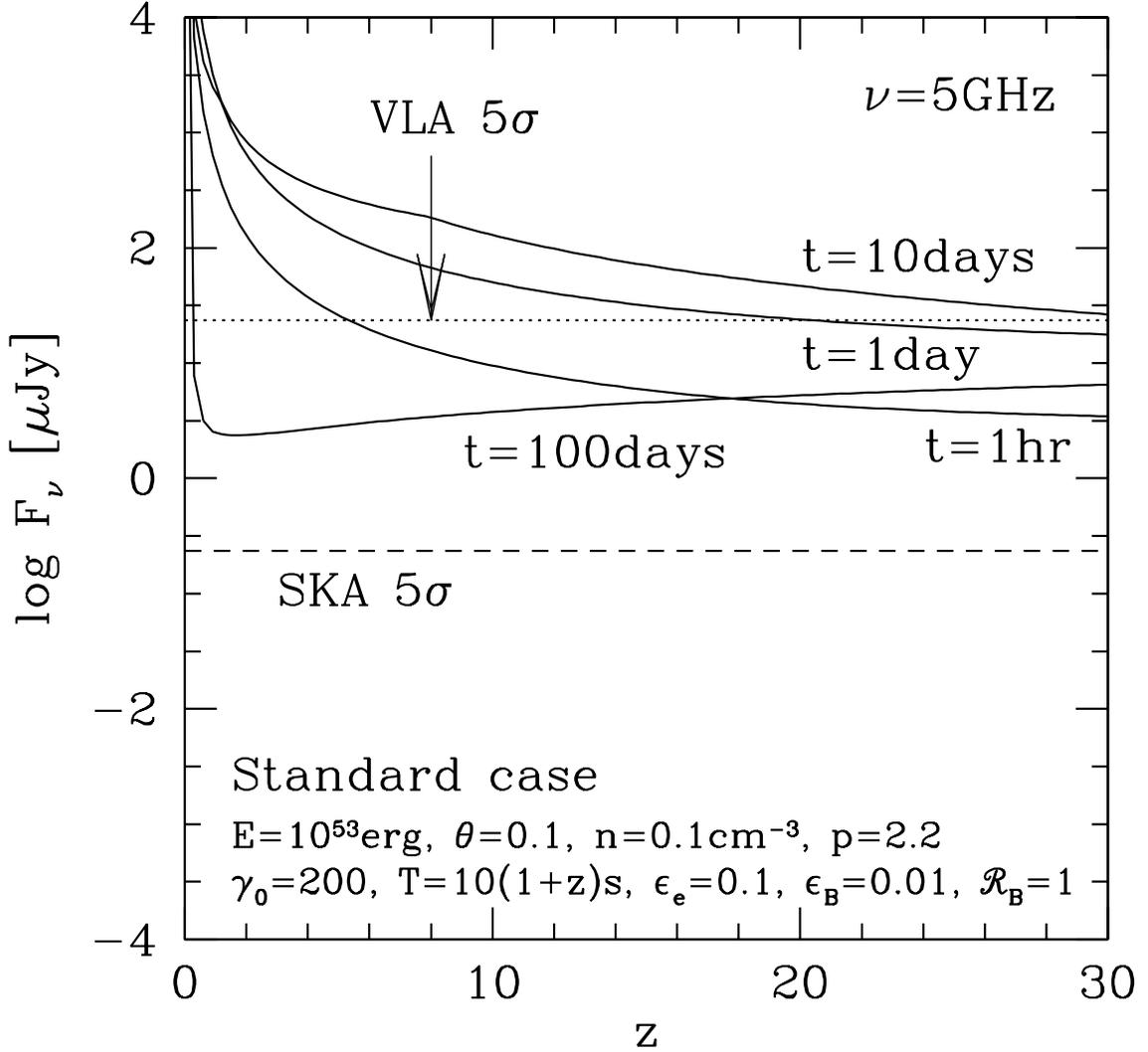}
\caption{\label{fig:standard:z}
GRB afterglow total flux from the forward and reverse shock 
emission at an observed frequency $\nu=5$ GHz
and observed times $1$ hr, $1$ day, $10$ days and $100$ days,
as a function of redshift $z$. 
The standard model parameters used are:
isotropic equivalent energy $E=10^{53}$ erg,
opening half-angle $\theta=0.1$,
ISM density $n=0.1$ cm$^{-3}$,
electron spectral index $p=2.2$,
initial Lorentz factor $\gamma_{0}=200$,
GRB duration in the source frame $T/(1+z)=10$ s,
and plasma parameters $\epsilon_{e}=0.1$,
$\epsilon_{B}=0.01$, ${\cal R}_{B}=1$.
The $5 \sigma$ sensitivities of the VLA (dotted line) and SKA (dashed line)
for an integration time $t_{\rm int}=1$ day and
a band width $\Delta \nu=50$ MHz are also shown.
}
\end{figure}

Figure~\ref{fig:standard:z} shows the total fluxes
of the forward and reverse shock emission
at the observed frequency $\nu=5$ GHz
and the observed times, $t=1$ hr, $1$ day, $10$ days and $100$ days,
as a function of redshift $z$.
The $5 \sigma$ sensitivities of the VLA and SKA are also shown.
We can see that the radio afterglows can be detected 
up to $z \sim 30$ even by the current VLA.
Here we use the sensitivity,
\beqa
F_{\nu}^{\rm sen}&=&\frac{{\rm SNR} \cdot 2 k T_{\rm sys}}
{A_{\rm eff} \sqrt{2 t_{\rm int} \Delta \nu}}
\nonumber\\
&\sim& 23 {\rm \mu Jy} 
\left(\frac{\rm SNR}{5}\right) 
\left(\frac{2 \times 10^{6} {\rm cm}^{2} {\rm K}^{-1}}
{A_{\rm eff}/T_{\rm sys}}\right)
\left(\frac{1 {\rm day}}{t_{\rm int}}\right)^{1/2}
\left(\frac{50 {\rm MHz}}{\Delta \nu}\right)^{1/2},
\label{eq:fsen}
\eeqa
assuming the signal-to-noise ratio (SNR) $5$,
the integration time $t_{\rm int}=1$ day,
the band width $\Delta \nu=50$ MHz,
$A_{\rm eff}/T_{\rm sys} \sim 2 \times 10^{6}$ cm$^{-2}$ K$^{-1}$
for the VLA,
and $A_{\rm eff}/T_{\rm sys} \sim 2 \times 10^{8}$ cm$^{-2}$ K$^{-1}$
for the SKA.
Note that LOFAR will has the frequency range 
from $\sim 10$ to $\sim 250$ MHz, and not at $\sim 5$ GHz.
In Figure~\ref{fig:standard:z},
the forward shock emission usually dominates the reverse shock one.
We can also see that the redshift dependence is rather weak.

\begin{figure}
\plotone{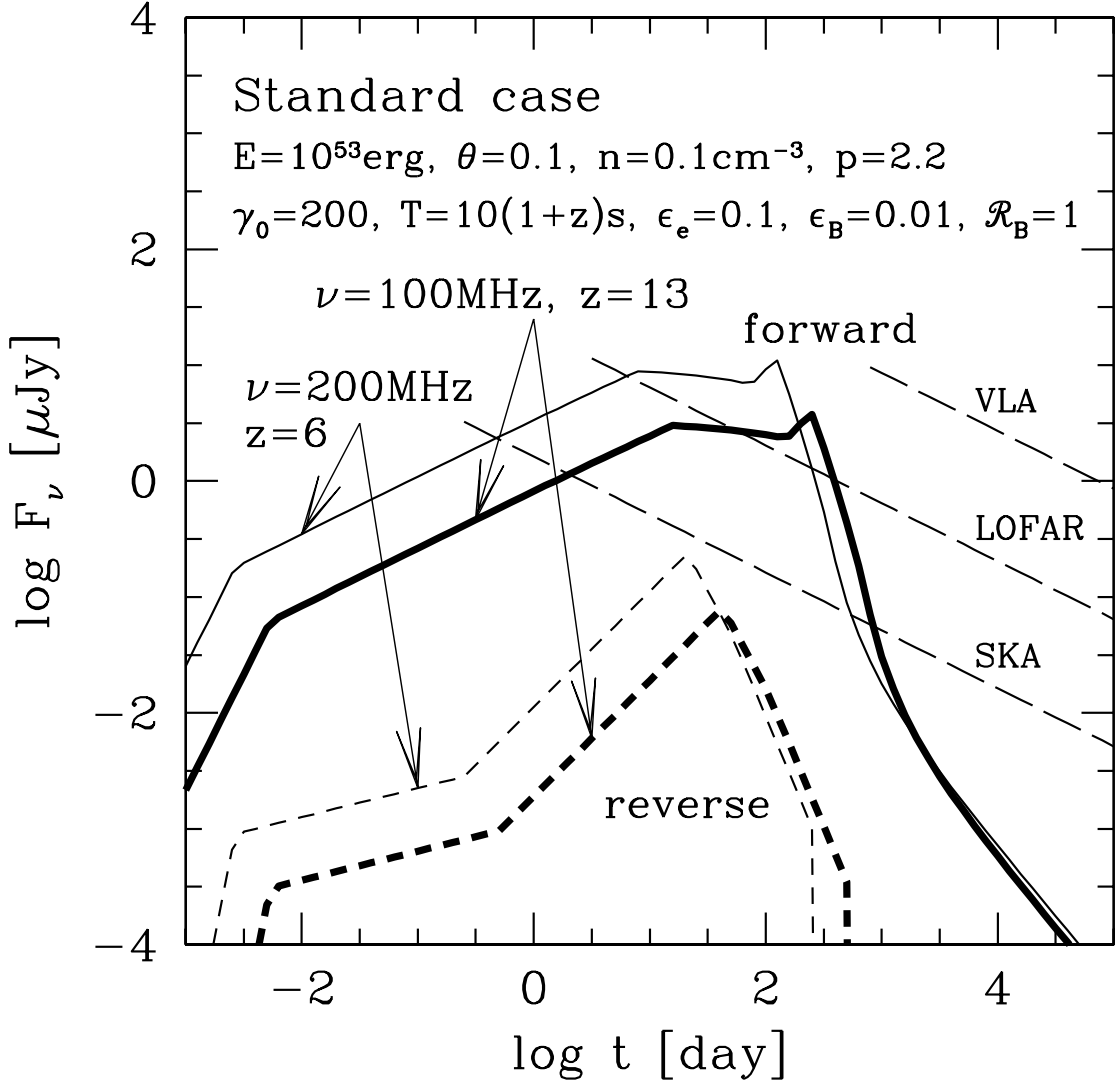}
\caption{\label{fig:standard}
GRB afterglow forward shock (solid line)
and reverse shock (dashed line) fluxes,
shown as a function of the observed time $t$
at frequencies near the redshifted 21 cm radiation 
for two representative redshifts.
We used the standard GRB model parameters:
isotropic equivalent energy $E=10^{53}$ erg,
opening half-angle $\theta=0.1$,
ISM density $n=0.1$ cm$^{-3}$,
electron spectral index $p=2.2$,
initial Lorentz factor $\gamma_{0}=200$,
GRB duration in the source frame $T/(1+z)=10$ s,
and plasma parameters $\epsilon_{e}=0.1$,
$\epsilon_{B}=0.01$, ${\cal R}_{B}=1$.
The thin lines are for a redshift $z=6$ and 
observed frequency $\nu=200$ MHz, while
thick lines are for $z=13$ and $\nu=100$ MHz.
The $5 \sigma$ sensitivities of the VLA, LOFAR and SKA
for integration times of one-third of $t$ and
band width $\Delta \nu=50$ MHz are also shown
by the long dashed lines.
}
\end{figure}

Figure~\ref{fig:standard} shows the light curves of the standard GRB
afterglows at $z=6$ and $\nu \sim 200$ MHz and at $z=13$ and 
$\nu \sim 100$ MHz, which correspond to the redshifted 21 cm band.
The $5 \sigma$ sensitivities of the VLA, LOFAR and SKA are also shown.
We see that LOFAR and SKA can detect the standard afterglows at 
low frequencies $\sim 100$ MHz and high redshift $z \sim 10$, 
but it may be difficult to detect them with the VLA.
Here we assumed an integration time of one-third of the observed 
time $t$, a band width $\Delta \nu=50$ MHz, 
$A_{\rm eff}/T_{\rm sys} \sim 3 \times 10^{5}$ cm$^{-2}$ K$^{-1}$
for the VLA,
$A_{\rm eff}/T_{\rm sys} \sim 4 \times 10^{6}$ cm$^{-2}$ K$^{-1}$
for the LOFAR,
and $A_{\rm eff}/T_{\rm sys} \sim 5 \times 10^{7}$ cm$^{-2}$ K$^{-1}$
for the SKA.
Note that $A_{\rm eff}/T_{\rm sys}$ at $\sim 100$ MHz is lower than 
that at $\sim 5$ GHz because of the galactic background.

From Figure~\ref{fig:standard} one sees that the peak flux is about 
$\sim 1$-$10$ $\mu$Jy at the redshifted frequency of the 21 cm line,
$\nu=1420/(1+z)$ MHz, for $z \simg 6$.  At these frequencies, the 
forward shock is brighter than the reverse shock, and the dependence 
on the redshift is rather weak. For most of the time the afterglow 
flux is below the self-absorption frequency $\nu_{a}$, so the flux 
is limited by the blackbody value in equation (\ref{eq:self}).
Since the flux is proportional to the electron temperature and the 
reverse shock has lower temperature than the forward shock, the 
reverse shock is dimmer than the forward shock for our parameters.
The first break in the forward shock emission at $t \sim 10^{-2}$ day
corresponds to the shock crossing time $t_{\times}$ in equation (\ref{eq:tcross}),
the second break at $t \sim 10$ day corresponds to
the jet break time $t_{\theta}$ in equation (\ref{eq:ttheta}),
the third break at $t \sim 10^{2}$ day is due to
the characteristic synchrotron frequency $\nu_{m,f}$ crossing
the observed frequency $\nu$,
the final peak is due to the self-absorption frequency $\nu_{a,f}$
crossing the observed frequency $\nu$,
and the non-relativistic phase begins soon after the final peak
(see \S~\ref{sec:model}).
The self-absorbed flux depends on the density as 
$F_{\nu,{\rm BB}} \propto n^{-1/2}$
at $t_{\times}<t<t_{\theta}$ if $\nu<\nu_{m,f}<\nu_{c,f}$
from equation (\ref{eq:self}).
Thus the flux is higher for a lower density.
However, if the density is too low, the synchrotron flux 
in equations (\ref{eq:slow}) and (\ref{eq:fast}) falls below the 
blackbody flux in equation (\ref{eq:self})
and the observed flux decreases as the density decreases, 
since $F_{\nu,{\rm max},f} \propto n^{1/2}$
at $t_{\times}<t<t_{\theta}$ if $\nu<\nu_{m,f}<\nu_{c,f}$.
We find that the flux is maximized at $n\sim 10^{-2}$ cm$^{-3}$
for our parameters (see also \S~\ref{sec:den}).

\subsection{Energetic GRB case}

The first generation of stars are expected to be metal-free
and very massive \citep{abel02,bromm02,omukai03}.
If the collapse of some of these population III stars leads 
to the launch of relativistic jets,
they could lead to an order of magnitude more energetic GRBs
than their low redshift counterparts
\citep[e.g.,][]{meszaros03,fryer01,heger03}.
Here we study a hypothesized energetic GRB case with 
isotropic equivalent energy $E=10^{54}$ erg,
fixing the other parameters as in the standard case.

\begin{figure}
\plotone{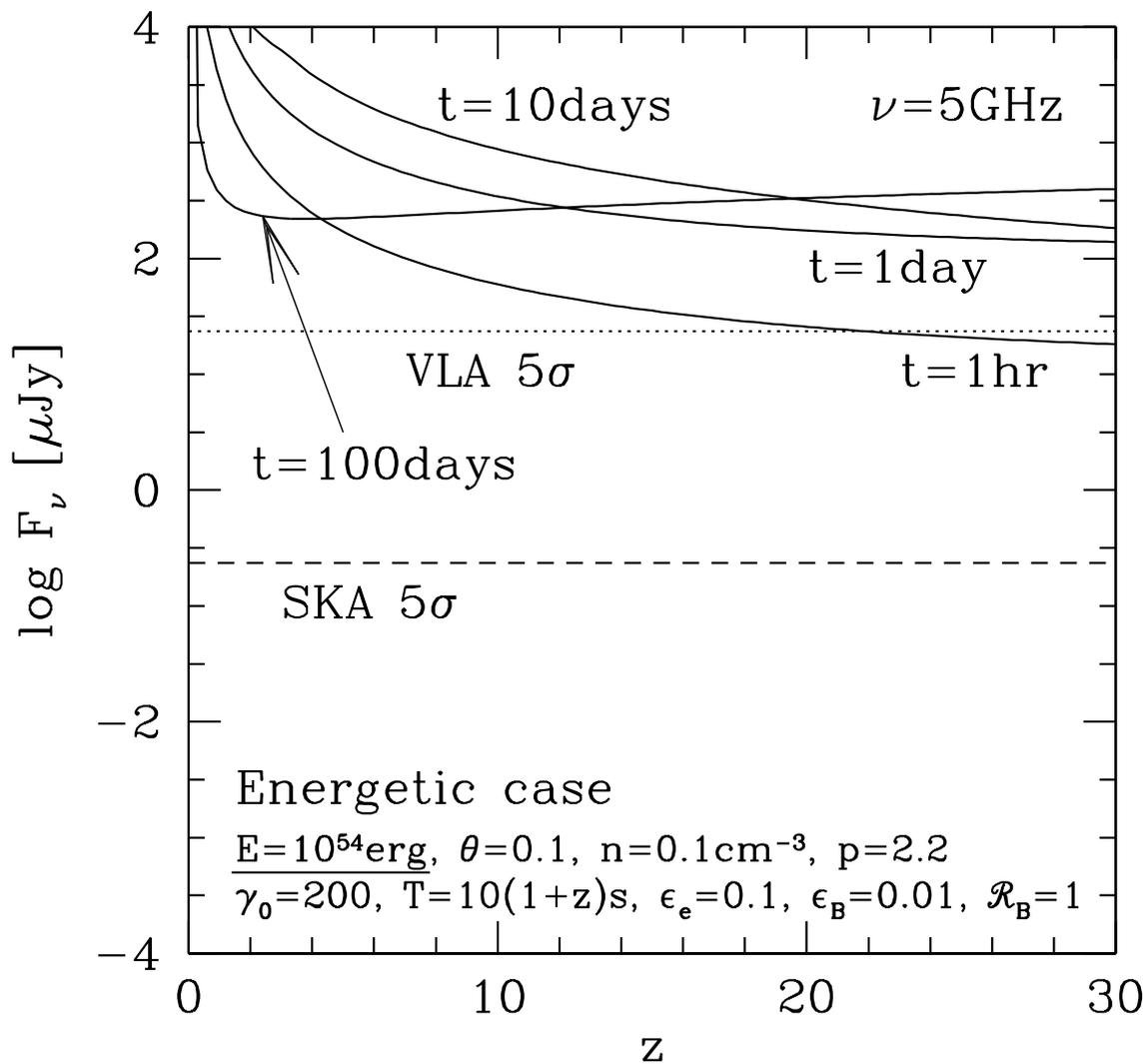}
\caption{\label{fig:ene:z}
GRB afterglow fluxes, same as in Figure~\ref{fig:standard:z} except for the
isotropic equivalent energy, which is taken here as $E=10^{54}$ erg.
}
\end{figure}

Figure~\ref{fig:ene:z} shows the total fluxes
of the forward and reverse shock emission
at the observed frequency $\nu=5$ GHz
and the observed times, $t=1$ hr, $1$ day, $10$ days and $100$ days,
as a function of redshift $z$,
together with the $5 \sigma$ sensitivities of the VLA and SKA.
By comparing Figures~\ref{fig:ene:z} and \ref{fig:standard:z}
we see that the detection is easier than in the standard case,
and even the VLA could easily detect such energetic GRB afterglows 
beyond $z \sim 30$.

Figure~\ref{fig:ene} shows the light curves of the energetic 
GRB afterglows at $z=6$ and $\nu \sim 200$ MHz, and at
$z=13$ and $\nu \sim 100$ MHz.
The $5 \sigma$ sensitivities of the VLA, LOFAR and SKA are also shown.
This indicates that LOFAR and SKA can detect such energetic 
afterglows at low frequencies $\sim 100$ MHz and high redshift 
$z \sim 10$, while VLA may detect them if observed for a long
enough time. The predicted peak flux is about $\sim 10$-$10^{2}$ $\mu$Jy
at the redshifted 21 cm frequency, $\nu=1420/(1+z)$ MHz, for $z \simg 6$.

\subsection{High density external medium case}\label{sec:den}

Even at high redshifts, the ambient density into which the
shell expands  could be low, $n\siml 1$ cm$^{-3}$ as assumed in 
the standard case (\S~\ref{sec:standard}), since the strong 
radiation from the progenitor star may evacuate a cavity in 
the surrounding gas \citep{whalen04,kitayama04}.
If this does not happen, however, the external density could be 
higher at high redshifts, based on hierarchical structure 
formation scenarios. The predicted evolution may be, e.g., 
$n \propto (1+z)^{4}$ for a fixed host galaxy mass \citep{ciardi00}.
Since little is known observationally about the ambient density at 
high redshifts, we assume a nominal high density case with 
$n=10^{2}$ cm$^{-2}$. Other parameters are the same as the standard case.

\begin{figure}
\plotone{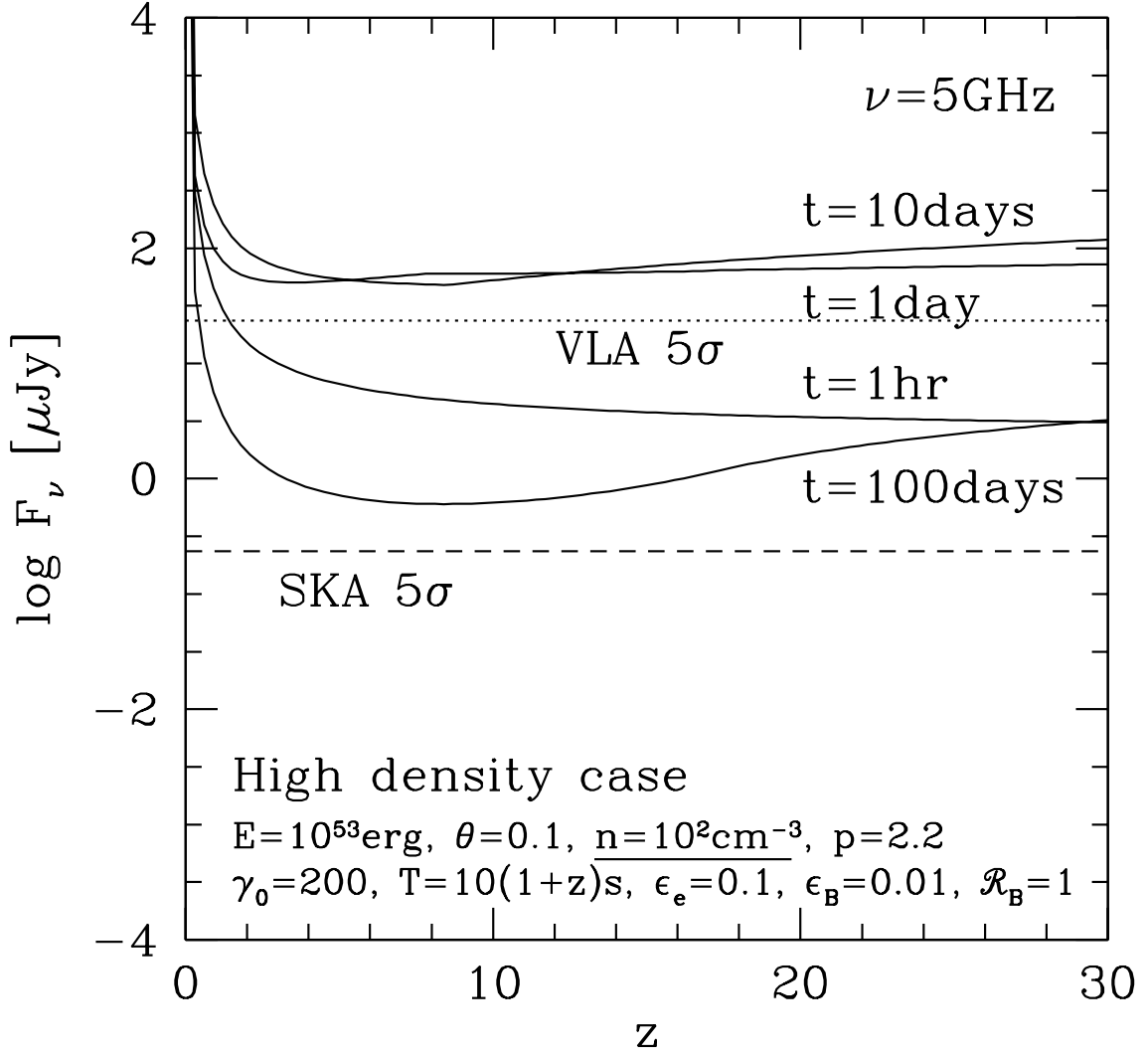}
\caption{\label{fig:high:z}
GRB afterglow fluxes as in Figure~\ref{fig:standard:z} except for the ISM 
density, which is taken here as $n=10^{2}$ cm$^{-3}$.
}
\end{figure}

Figure~\ref{fig:high:z} shows the total fluxes
of the forward and reverse shock emission
at the observed frequency $\nu=5$ GHz
and the observed times $t=1$ hr, $1$ day, $10$ days and $100$ days,
as a function of redshift $z$,
together with the $5 \sigma$ sensitivities of the VLA and SKA.
By comparing Figures~\ref{fig:high:z} and \ref{fig:standard:z}
we can see that the detection at this frequency is easier than in 
the standard case, and even the VLA can easily detect such high 
density GRB afterglows beyond $z \sim 30$.

\begin{figure}
\plotone{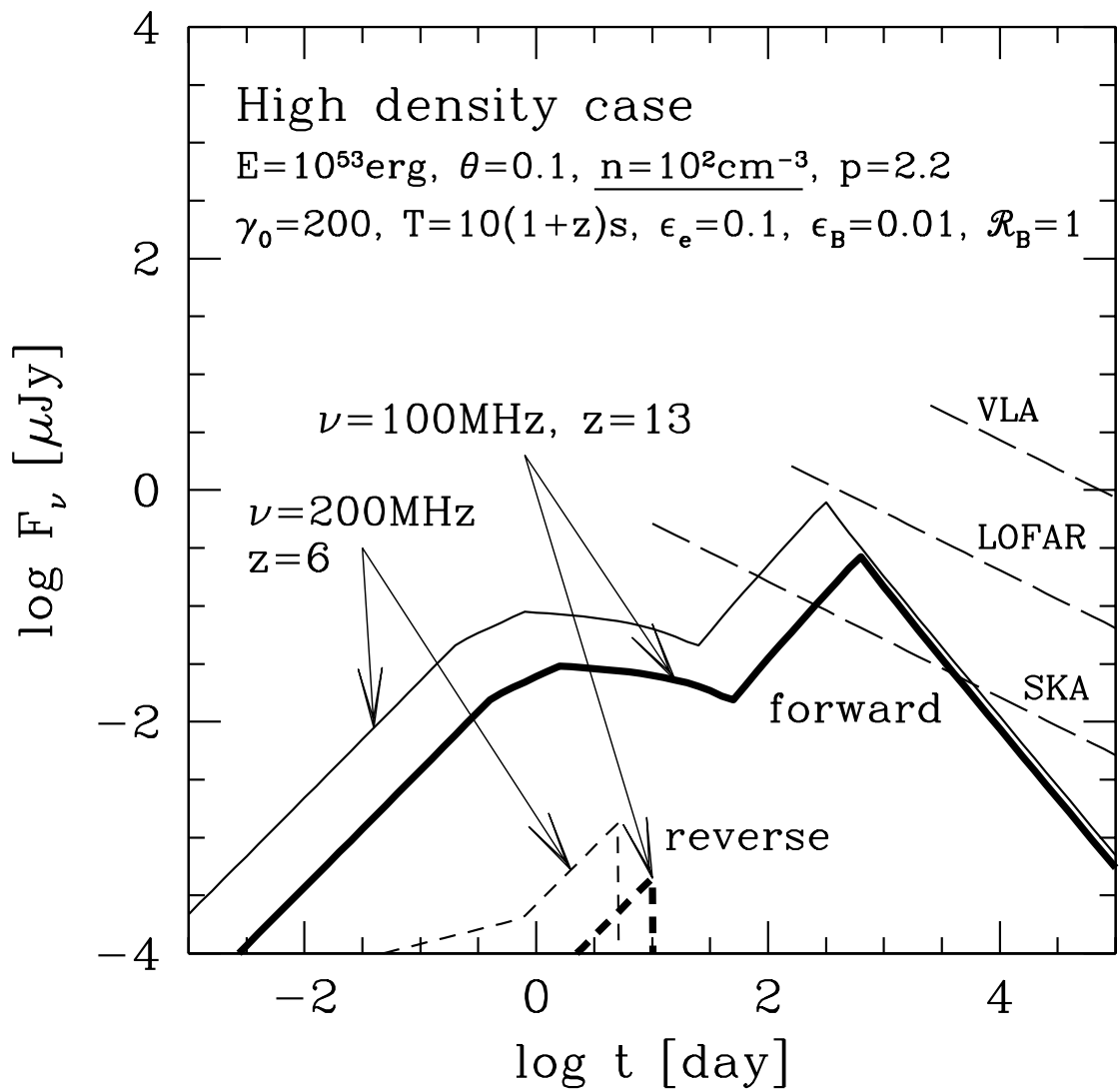}
\caption{\label{fig:high}
GRB afterglow fluxes at two redshifted 21 cm frequencies,
as in Figure~\ref{fig:standard} except for the ISM density,
which is taken here as $n=10^{2}$ cm$^{-3}$.
}
\end{figure}

Figure~\ref{fig:high} shows the light curves
of the high density GRB afterglows at $z=6$ and $\nu \sim 200$ MHz and at
$z=13$ and $\nu \sim 100$ MHz.
The $5 \sigma$ sensitivities of the VLA, LOFAR and SKA are also shown.
This shows that the detection of the high density GRB afterglows at 
low frequencies $\sim 100$ MHz and high redshift $z \sim 10$ is 
difficult. Even the SKA would require very long observing times.
The flux is dimmer than in the standard case in 
Figure~\ref{fig:standard}, because the self-absorption strongly 
suppresses the low frequency fluxes in this high density case.
From Figure~\ref{fig:high}, we see that the peak flux is about 
$\sim 10^{-1}$-$1$ $\mu$Jy at the redshifted 21 cm frequency,
$\nu=1420/(1+z)$ MHz, for $z \simg 6$.

\subsection{Long duration GRB case}

The GRB duration at high redshifts could in principle be 
much longer than at low redshifts, since more massive 
progenitor stars could lead to longer accretion times and longer 
jet feeding times \citep{fryer01}. A long duration may also 
be due to continued fallback of material when the explosion is 
too mild \citep{macfadyen01}.  BATSE did detect a few bursts 
longer than $1000$ s, with its standard peak-flux trigger, 
instead of a fluence trigger. Swift has the ability to operate
with a fluence trigger, which potentially could lead to the
detection of longer GRBs. As an example of long duration GRBs,
we investigate a case whose duration in the source frame is
$T/(1+z)=1000$ s. Other parameters are the same as in the standard case.

\begin{figure}
\plotone{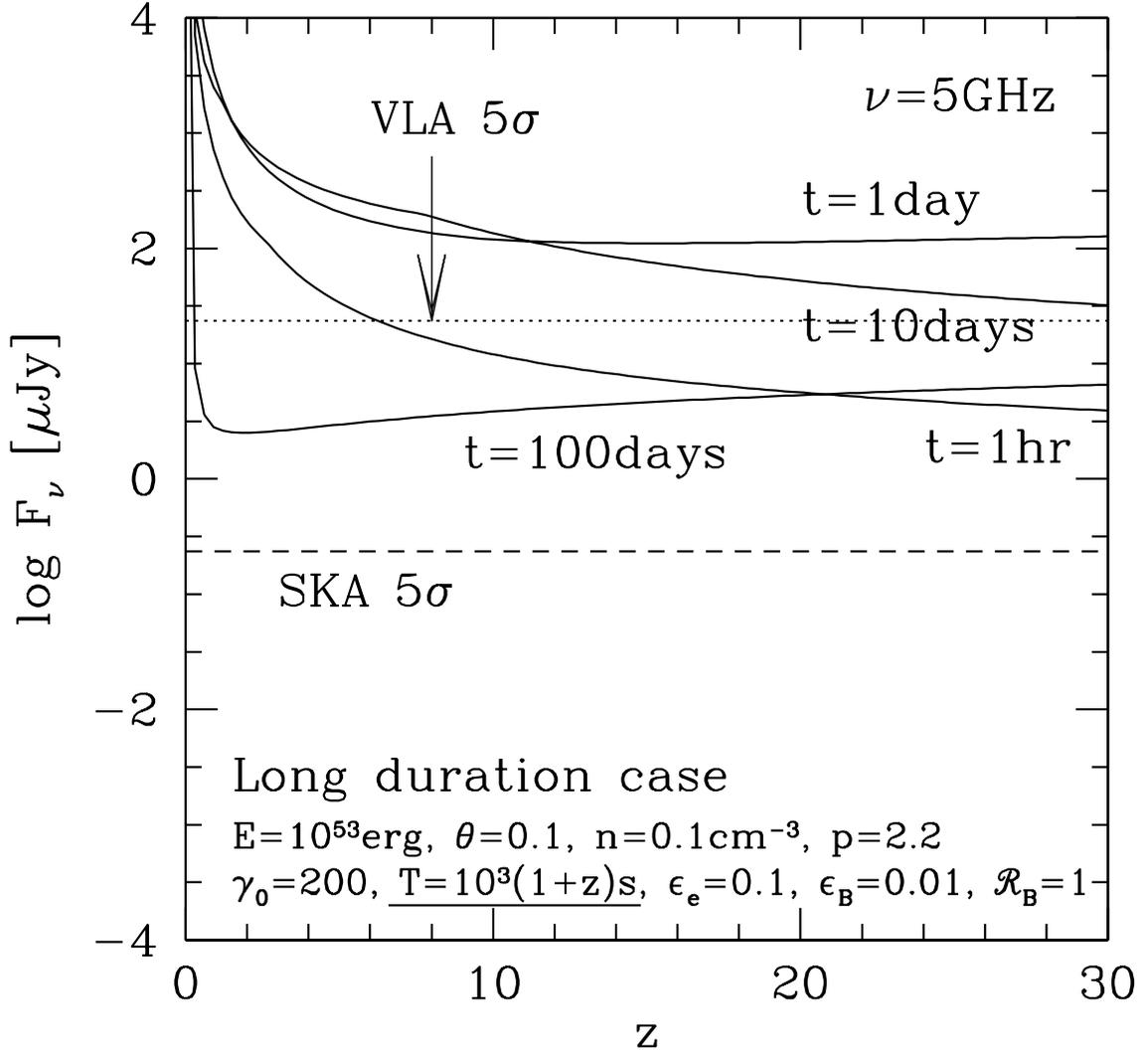}
\caption{\label{fig:long:z}
GRB afterglow fluxes as in Figure~\ref{fig:standard:z} except for the 
GRB duration in the source frame,
which is taken here as $T/(1+z)=10^{3}$ s.
}
\end{figure}

Figure~\ref{fig:long:z} shows the total fluxes
of the forward and reverse shock emission
at an observed frequency $\nu=5$ GHz
and at observed times, $t=1$ hr, $1$ day, $10$ days and $100$ days,
as a function of redshift $z$,
together with the $5 \sigma$ sensitivities of the VLA and SKA.
Comparing Figures~\ref{fig:long:z} and \ref{fig:standard:z}
we see that the detection of long bursts is easier than for the
standard case of duration $T/(1+z)=10$ s, and even the VLA can 
easily detect such long burst afterglows beyond $z \sim 30$.
This is because the reverse shock emission at early times 
$t \siml 10$ day is enhanced due to the long duration.

\begin{figure}
\plotone{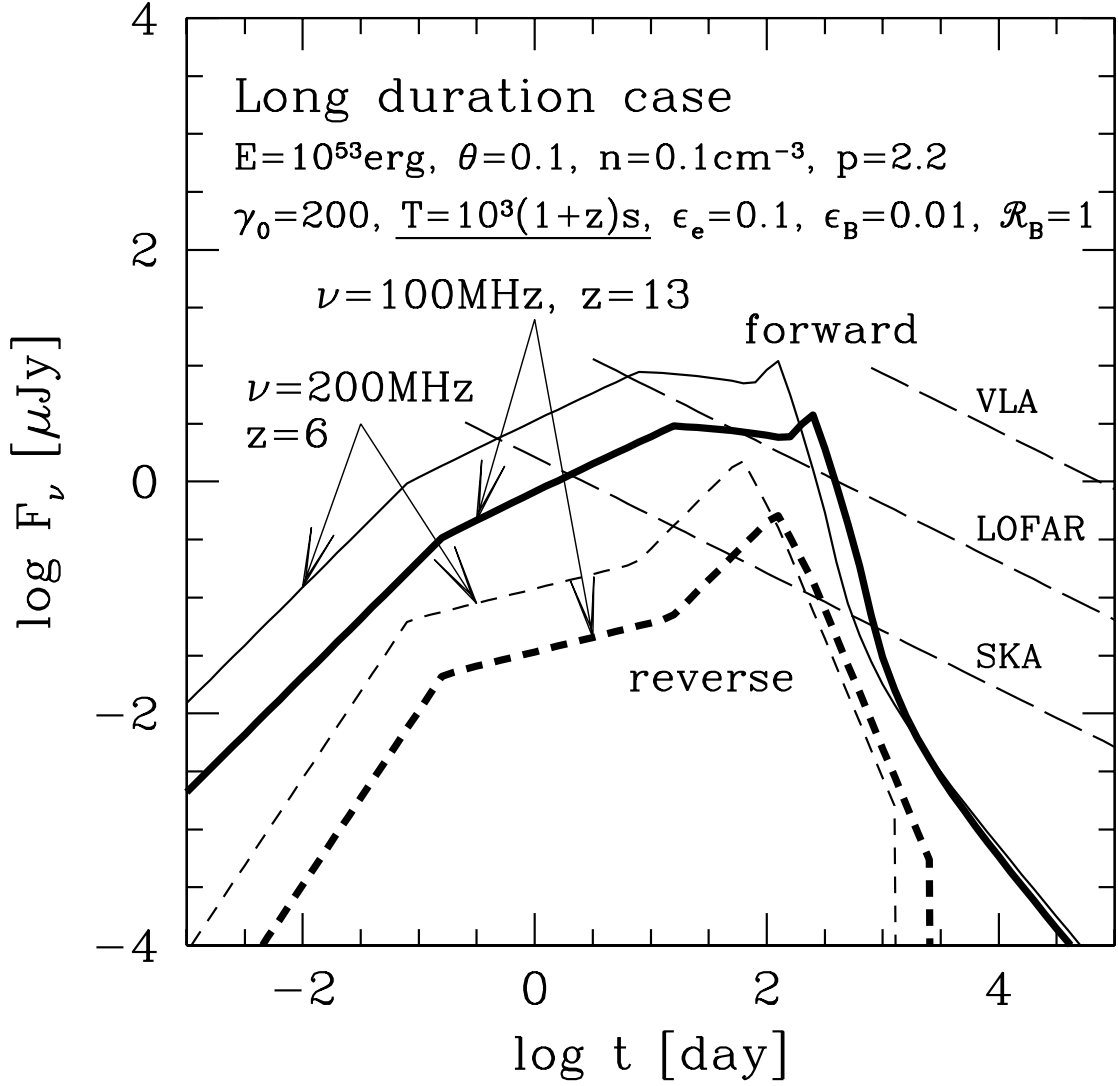}
\caption{\label{fig:long}
GRB afterglow fluxes at two redshifted 21 cm frequencies,
same as in Figure~\ref{fig:standard} except for the GRB duration
in the source frame, which is taken here as $T/(1+z)=10^{3}$ s.
}
\end{figure}

Figure~\ref{fig:long} shows the long duration GRB afterglow
light curves at $z=6$ and $\nu \sim 200$ MHz and at
$z=13$ and $\nu \sim 100$ MHz.
The $5 \sigma$ sensitivities of the VLA, LOFAR and SKA are also shown.
Although the reverse shock emission is brighter than in the standard 
case in Figure~\ref{fig:standard},
it is still below the forward shock emission
at low frequencies $\sim 100$ MHz.
Since the forward shock emission is similar to the standard case,
the detectability at these frequencies is similar to the standard 
case.  LOFAR and SKA should be able to detect the long duration 
GRB afterglows at low frequencies $\sim 100$ MHz and high redshift 
$z \sim 10$, while it appears difficult for the VLA to detect them.
As seen in Figure~\ref{fig:long}, the peak flux is about 
$\sim 1$-$10$ $\mu$Jy at the redshifted 21 cm frequency,
$\nu=1420/(1+z)$ MHz, for $z \simg 6$.

\subsection{Magnetized GRB ejecta case}

Several observations suggest that the ejected shell may be
strongly magnetized.
A large linear polarization in the gamma-ray emission of GRB 021206
has been interpreted as implying a uniform ordered magnetic field 
over the visible region \citep{coburn03}, which could be a magnetic 
field advected from the central source 
\citep[see however][]{rutledge04,wigger04}.
The early afterglows of GRB 990123 and GRB 021211 also may imply 
that the reverse shock region has a stronger magnetic field
than the forward shock region \citep{zhang03b,fan02b}.
Although further studies are necessary to verify these suggestions 
\citep[e.g.,][]{zhang04,matsumiya03,fan04}, it is useful to examine as
a possibility the strongly magnetized case.  Here we adopt a ratio 
of reverse to forward magnetic field strengths ${\cal R}_{B}=5$,
with the other parameters the same as in the standard case.

\begin{figure}
\plotone{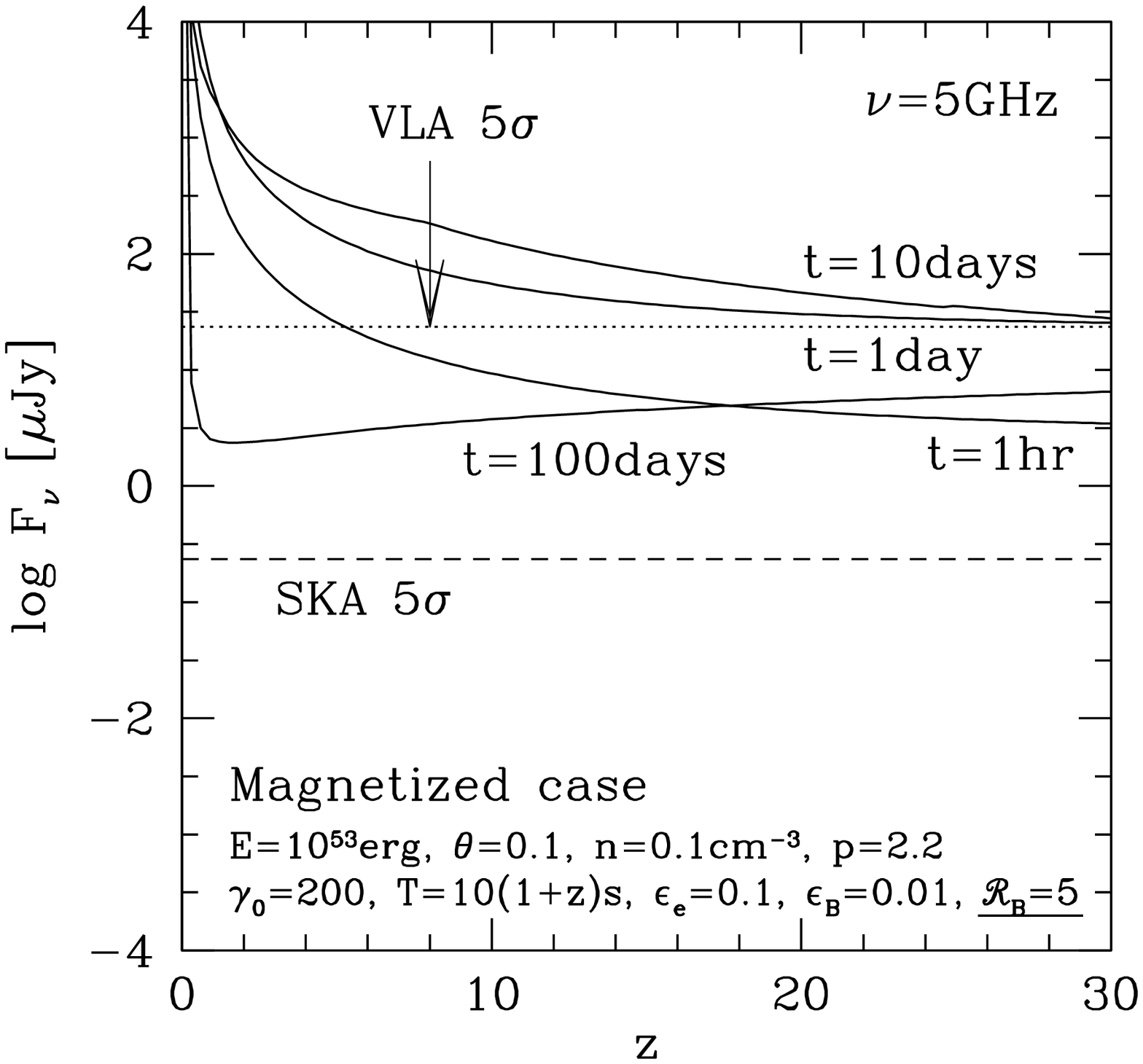}
\caption{\label{fig:magnetized:z}
GRB afterglow fluxes, as in Figure~\ref{fig:standard:z} 
except for the magnetic field
in the reverse shock, which is taken here as 
${\cal R}_{B}=(\epsilon_{B,r}/\epsilon_{B,f})^{1/2}=5$.
}
\end{figure}

Figure~\ref{fig:magnetized:z} shows the total fluxes
of the forward and reverse shock emission
at the observed frequency $\nu=5$ GHz
for observed times $t=1$ hr, $1$ day, $10$ days and $100$ days,
as a function of redshift $z$,
together with the $5 \sigma$ sensitivities of the VLA and SKA.
Comparing Figures~\ref{fig:magnetized:z} and \ref{fig:standard:z}
we see that the detection of the magnetized GRB afterglow is 
somewhat easier than for the standard case, and even the current 
VLA could detect such magnetized GRB afterglows up to $z \sim 30$.
A small increase in the flux is caused by the reverse shock 
emission at early times $t \siml 10$ day, due to the higher reverse
magnetic field.

\begin{figure}
\plotone{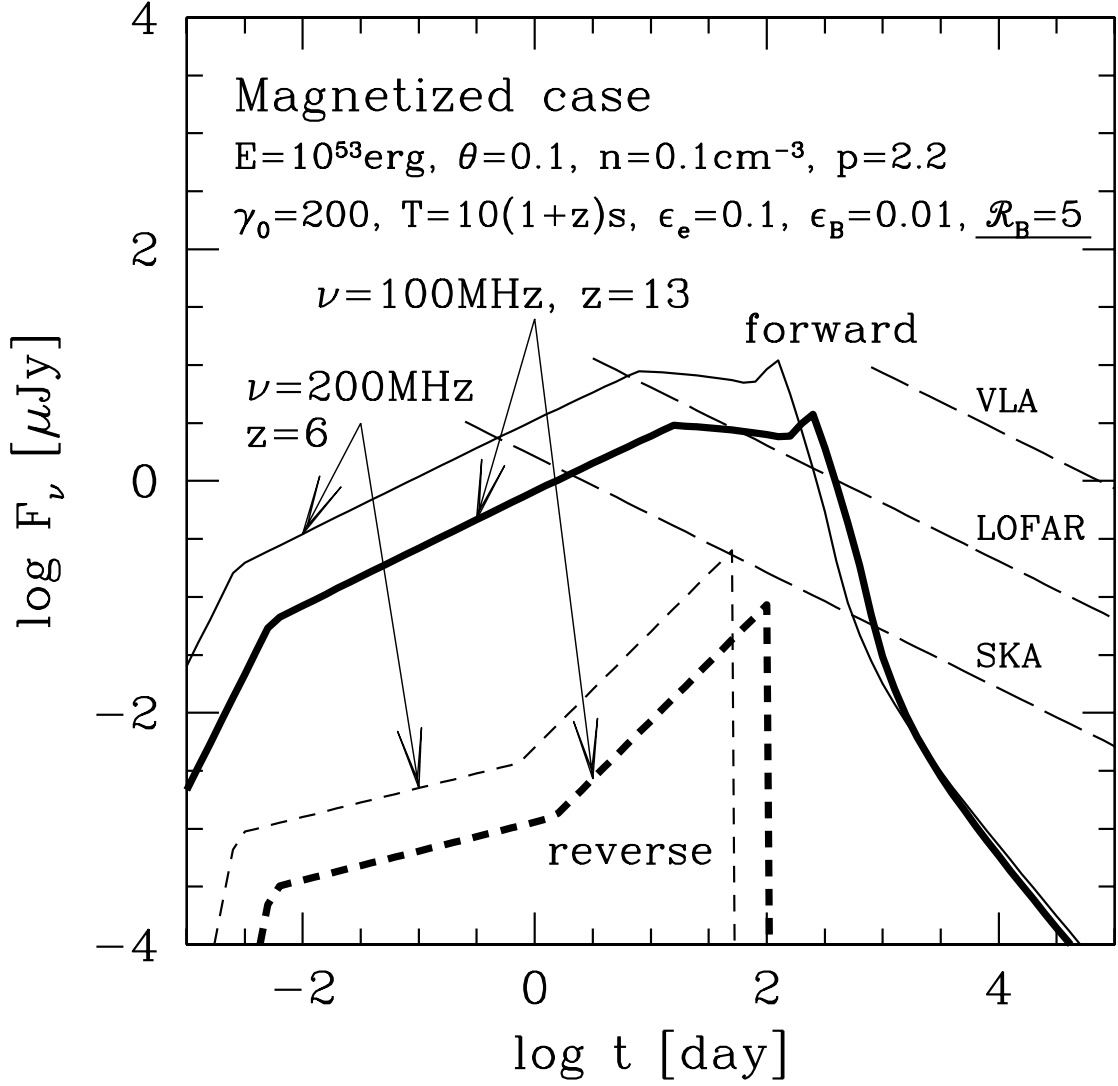}
\caption{\label{fig:magnetized}
GRB afterglow fluxes at two redshifted 21 cm frequencies,
as in Figure~\ref{fig:standard} except for the magnetic field
in the reverse shock,
which is taken here as ${\cal R}_{B}=(\epsilon_{B,r}/\epsilon_{B,f})^{1/2}=5$.
}
\end{figure}

Figure~\ref{fig:magnetized} shows the light curves of the magnetized 
GRB afterglows at $z=6$ and $\nu \sim 200$ MHz and at
$z=13$ and $\nu \sim 100$ MHz.
The $5 \sigma$ sensitivities of the VLA, LOFAR and SKA are also shown.
The reverse shock emission at low frequencies $\sim 100$ MHz
is still below the forward shock emission, being dimmer at these
frequencies than in the standard case in Figure~\ref{fig:standard}.
Since the forward shock emission is similar to the standard case,
the detectability is similar to that in the standard case.
We see that LOFAR and SKA can detect the magnetized afterglows 
at low frequencies $\sim 100$ MHz
and high redshift $z \sim 10$, while
the VLA may have difficulty detecting them.
From Figure~\ref{fig:magnetized},
the peak flux is seen to be about $\sim 1$-$10$ $\mu$Jy
at the redshifted 21 cm frequency,
$\nu=1420/(1+z)$ MHz, for $z \simg 6$.

\subsection{Hypernova radio afterglow}

The association of Type Ib/c supernovae (SNe) with GRBs
has confirmed the massive stellar origin of long GRBs
\citep[e.g.,][]{zhang03,meszaros02}.
Only a small fraction of core collapse SNe
appear to lead to GRB jets with high Lorentz factor 
$\gamma_{0} \simg 100$ \citep{berger03,soderberg04}.
However, there may be a larger number of SNe accompanied
by mildly relativistic ejecta, which could in principle 
involve more  energy than typical GRB jets \citep{macfadyen01,granot04}.
A large kinetic energy $\sim 10^{52}$ erg is also suggested in 
some core collapse SNe, by the optical light curve and spectral 
fits \citep{nomoto03,pod04}, and these objects appear in a  number
of cases not to be associated with detected GRB.
Such energetic supernovae (or hypernovae, HNe) may be powered by 
jets that fail to bore their way through the stellar envelope
\citep{khokhlov99,meszaros01}.
At high redshifts, stars are expected to be very massive 
\citep{abel02,bromm02,omukai03}, and it may be more frequent for 
the jet to be choked even while the jet could be $\sim 10$ times 
more energetic \citep{heger03,fryer01}.

Based on these considerations, an isotropic equivalent energy of 
$E\sim 10^{53}$ erg may be reasonable for high redshift HNe,
and depending on the energy conversion efficiency,
even an isotropic equivalent energy of $E\sim 10^{54}$ erg
may not be implausible for population III HNe \citep{fryer01,heger03}.
Keeping in mind the hypothetical nature of this assumption,
we examine an extreme HNe afterglow case with
isotropic equivalent energy $E=10^{54}$ erg,
ejecta opening half-angle $\theta=1/\sqrt{2}$
(nearly spherical)
and initial Lorentz factor $\gamma_{0}=2$
(mildly relativistic).
Other parameters are the same as in the standard GRB case.
We note that different authors define HNe differently. Here,
our definition of HNe is SNe with a large energy and a
mildly relativistic ejecta.

\begin{figure}
\plotone{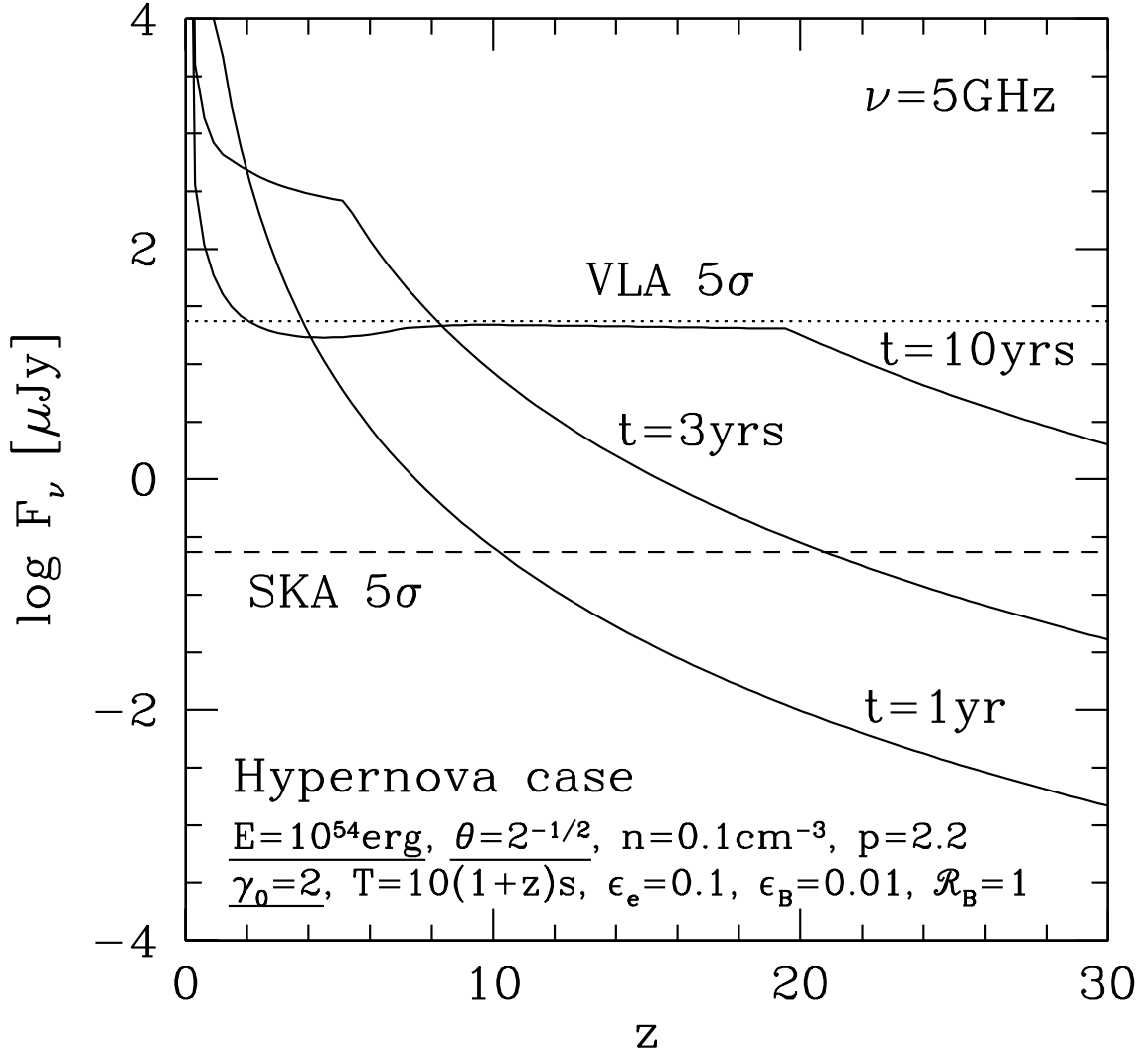}
\caption{\label{fig:HN:z}
Hypernova (HN) remnant total flux from the forward and reverse shock 
at the observed frequency $\nu=5$ GHz
and observed times $1$ yr, $3$ yrs and $10$ yrs,
shown as a function of redshift $z$. 
The model parameters are as for the standard GRB case
in Figure~\ref{fig:standard:z}, except for
the isotropic equivalent energy $E=10^{54}$ erg,
the opening half-angle $\theta=2^{-1/2}$
(nearly spherical)
and the initial Lorentz factor $\gamma_{0}=2$
(mildly relativistic).
The $5 \sigma$ sensitivities of the VLA (dotted line) and SKA (dashed line)
for an integration time $t_{\rm int}=1$ day and
a band width $\Delta \nu=50$ MHz are also shown.
}
\end{figure}

Figure~\ref{fig:HN:z} shows the total fluxes
of the forward and reverse shock emission
at the observed frequency $\nu=5$ GHz
and observed times $t=1$ yr, $3$ yrs and $10$ yrs,
as a function of redshift $z$,
together with the $5 \sigma$ sensitivities of the VLA and SKA.
This shows that the current VLA may be able to detect such
energetic HN afterglows up to $z \sim 20$ with $\sim 1$ day 
integration times, while SKA can easily detect them beyond $z \sim 30$.
We have also calculated a more moderate energy HN case with
$E \sim 10^{53}$ erg, as well as a high external density HN case 
with $n \sim 10^{2}$ cm$^{-3}$. The results for these cases (not
shown) indicate that the VLA can detect them out to $z \sim 8$ 
and beyond $z \sim 30$, respectively, while the SKA can easily 
detect them beyond $z \sim 30$.

\begin{figure}
\plotone{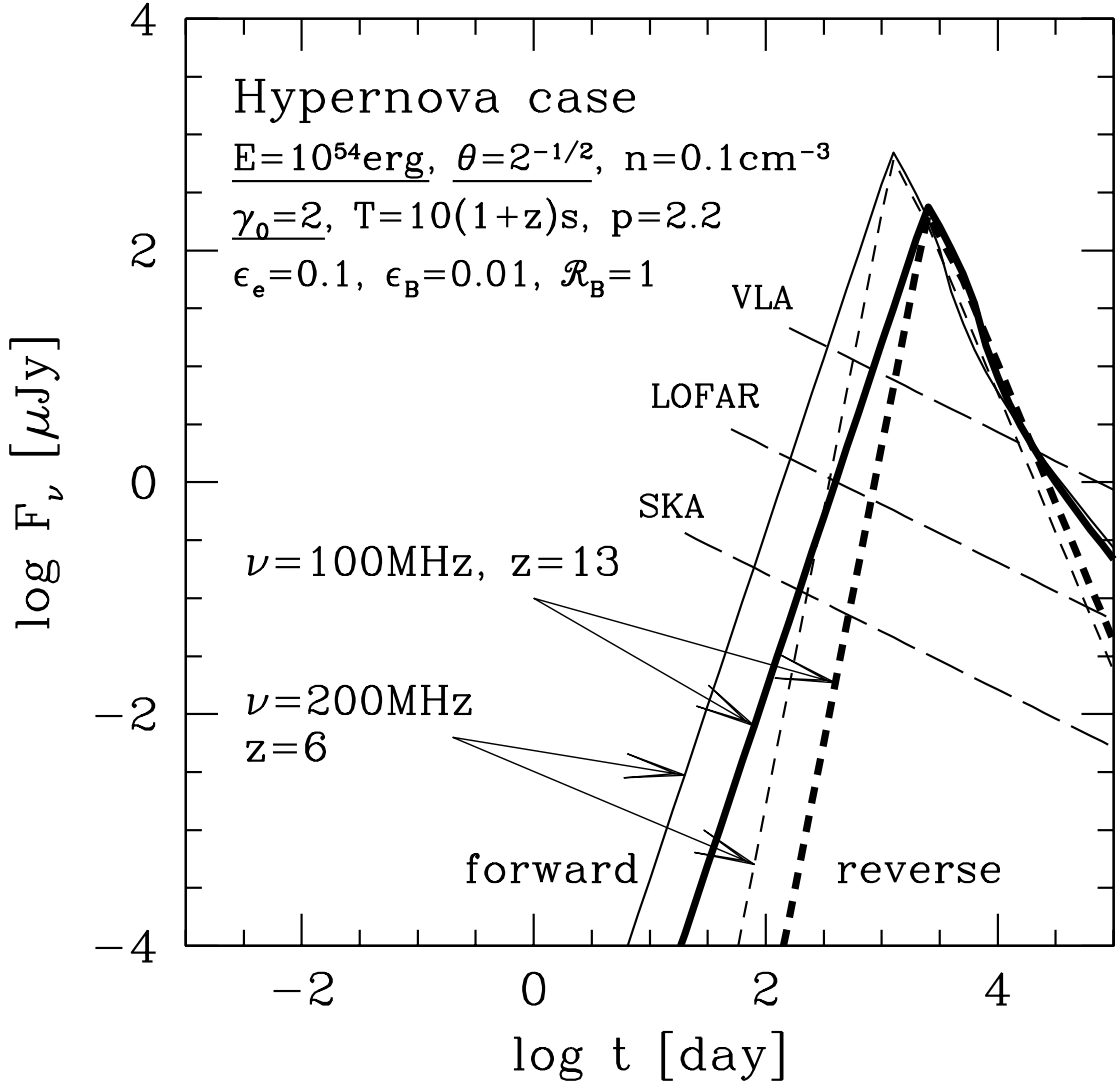}
\caption{\label{fig:HN}
Hypernova (HN) remnant fluxes at two redshifted 21 cm frequencies,
same as Figure~\ref{fig:standard} except for 
the isotropic equivalent energy $E=10^{54}$ erg,
the opening half-angle $\theta=2^{-1/2}$
(nearly spherical)
and the initial Lorentz factor $\gamma_{0}=2$
(mildly relativistic).
The $5 \sigma$ sensitivities of the VLA, LOFAR and SKA
for the integration time of one-third of $t$ and
the band width $\Delta \nu=50$ MHz are also shown
by the long dashed lines.
}
\end{figure}

Figure~\ref{fig:HN} shows the light curves of the energetic 
HN afterglows at $z=6$ and $\nu \sim 200$ MHz and at
$z=13$ and $\nu \sim 100$ MHz, as well as the
$5 \sigma$ sensitivities of the VLA, LOFAR and SKA.
This shows that the VLA, LOFAR and SKA can detect
such HN afterglows at low frequencies $\sim 100$ MHz
and high redshifts $z \sim 10$. The peak flux indicated 
is about $\sim 10^{2}$-$10^{3}$ $\mu$Jy at the redshifted 
21 cm frequency, $\nu=1420/(1+z)$ MHz, for $z \simg 6$.
The reverse shock emission is comparable to the forward shock 
emission, and the redshift dependence is weak.
We also calculated the moderate energy HN case $E \sim 10^{53}$ erg
and the high density HN case $n \sim 10^{2}$ cm$^{-3}$
to find that the peak flux at $\sim 100$ MHz in both cases
is about an order of magnitude below the flux 
in Figure~\ref{fig:HN}.

One caveat about the HN emission is that one may have no prior 
information about the sky position because of the preponderant
lack of detected gamma-ray emission.  However the number of HN 
afterglows at $z \simg 6$ on the sky is estimated as
$\sim 10^{3}$ events yr$^{-1} \times t_{\rm peak} 
f_{z} f_{\rm HN/GRB} \sim 10^{3}$,
where $t_{\rm peak} \sim 10^{3}$ days is the peak time of the HNe,
and we assume a fraction $f_{z} \sim 50$ \%
of all GRBs on the sky originating at $z \simg 6$ \citep{bromm02a}
and a ratio of the HN rate to the GRB one $f_{\rm HN/GRB}\sim 1$.
Thus the imaging of $\sim 40$ deg$^{2}$ of sky would lead to
one HN afterglow at $z \simg 6$, while the field of view of the SKA
may be larger than this. Thus, HN afterglows could be detected 
as a by-product of other related or unrelated observing programs.

It is also interesting to note that the HN afterglows at high 
redshift may be brighter than their low redshift counterparts,
since the energy may be larger at high redshifts and the redshift 
dependence of the flux is weak when the other parameters are fixed.
Thus the orphan afterglow searches \citep[e.g.,][]{totani02}
may preferentially find the high redshift afterglows,
although the discrimination from variable AGNs
may be difficult \citep{levinson02}.

\section{21 cm absorption at $z \simg 6$}\label{sec:z>6}

At redshifts $z \simg 6$ one may expect 21 cm absorption
of the radio afterglow emission by the diffuse neutral component
of the intergalactic medium (IGM) \citep{carilli02},
as well as by collapsed gas clouds, such as in
minihalos and galactic disks \citep{furlanetto02}.
The 21 cm absorption by the IGM would provide constraints on
the reionization history of the universe \citep{carilli02}, 
while the 21 cm forest due to gas clouds would provide information 
about the structure and galaxy formation at high redshifts.
The 21 cm absorption line due to the host galaxy may provide
GRB and host galaxy redshifts \citep{furlanetto02}, independently 
of other methods such as the Ly$\alpha$ cutoff \citep{lamb00} 
and X-ray lines \citep{meszaros03,gou04b}.

We consider first the optical depth of the diffuse 
IGM to the 21 cm absorption, which is given by
\beqa
\tau=\frac{3 c^{3} h_{\rm P} A_{10} n_{\rm H_I}(z)}
{32 \pi k \nu_{0}^{2} T_{S} H(z)}
=0.010 \frac{T_{\rm CMB}}{T_{S}} \left(\frac{1+z}{10}\right)^{1/2}
x_{\rm H_I},
\label{eq:tau}
\eeqa
where $T_{\rm CMB}=2.73 (1+z)$ K is the CMB temperature at redshift $z$,
$T_{S}$ is the spin temperature of the IGM,
$x_{\rm H_I}$ is the neutral hydrogen fraction,
$h_{\rm P}$ is Plank's constant,
$k$ is Boltzmann's constant,
$A_{10}=2.85 \times 10^{-15}$ s$^{-1}$ is the spontaneous emission 
coefficient of the 21 cm line,
$\nu_{0}=1420.4$ MHz is the source frame 21 cm frequency, 
$H(z) \simeq H_{0} \Omega_{m}^{1/2} (1+z)^{3/2}$
is the Hubble parameter at high $z$,
$n_{\rm H_{I}}$ is the neutral hydrogen density at $z$,
and we assume the helium mass fraction to be $24\%$
\citep{field59a,madau97}.

The main uncertainty in equation (\ref{eq:tau})
comes from the hydrogen spin temperature $T_{S}$.
The spin temperature is determined by the gas kinetic temperature
$T_{K}$ and the coupling of $T_{S}$ with $T_{K}$ and $T_{\rm CMB}$
\citep{madau97,tozzi00}.
In the presence of only the CMB, we have $T_{S}\sim T_{\rm CMB}$.
Once the early generation of stars starts emitting Ly$\alpha$ photons,
the Ly$\alpha$ pumping may tightly couple the spin temperature with 
the gas kinetic temperature well before the epoch of full reionization 
\citep[e.g.,][]{ciardi03}.
The kinetic temperature drops below the CMB temperature
without any heating because of the adiabatic cooling
due to the expansion of the universe,
while the kinetic temperature may be quickly increased above
the CMB temperature by the Ly$\alpha$ and X-ray heating
due to the first stars and the shock heating in the high density regions
\citep{carilli02,chen04}.
The uncertainty in the spin temperature is at least
$0.5 T_{\rm CMB} \siml T_{S} \siml 4 T_{\rm CMB}$ \citep{chen04}.
If $T_{S} \sim 4 T_{\rm CMB}$,
the optical depth is about $\tau \sim 0.0021 x_{\rm H_{I}}$
at $z \sim 6$ and $\tau \sim 0.0030 x_{\rm H_{I}}$ at $z \sim 13$
from equation (\ref{eq:tau}).

In order to detect the 21 cm absorption line,
the source flux should be larger than
\beqa
F_{\nu}^{\min} &=& 
\frac{2 k T_{\rm sys}}{A_{\rm eff} \sqrt{2 t \Delta \nu}}
\frac{\rm SNR}{\tau}
\nonumber\\
&\sim& 1.1 {\rm mJy} 
\left(\frac{\rm SNR}{5}\right) 
\left(\frac{0.002}{\tau}\right)
\left(\frac{5\times 10^{7} {\rm cm}^{2} {\rm K}^{-1}}
{A_{\rm eff}/T_{\rm sys}}\right)
\left(\frac{10 {\rm days}}{t_{\rm int}}\right)^{1/2}
\left(\frac{1 {\rm MHz}}{\Delta \nu}\right)^{1/2},
\label{eq:fmin}
\eeqa
where we assume a band width $\Delta \nu \sim 1$ MHz,
an integration time $t_{\rm int} \sim 10$ days,
the SNR 5,
the capability of the SKA
$A_{\rm eff}/T_{\rm sys}\sim 5\times 10^{7}$ cm$^{2}$ K$^{-1}$
at $\nu \sim 200$ MHz,
and the small $\tau$ limit.
Therefore $\sim 1$ mJy sources are needed
for the detection of the 21 cm absorption due to the diffuse IGM
if $T_{S} \sim 4 T_{\rm CMB}$.
If there was an epoch with low spin temperature 
$T_{S} \sim 0.5 T_{\rm CMB}$ \citep{chen04},
the required flux could be as low as $\sim 0.1$ mJy,
since $\tau \propto T_{S}^{-1}$
from equation (\ref{eq:tau}).
Note that the redshift is determined with the precision,
\beqa
\Delta z \sim \frac{\Delta \nu (1+z)^{2}}{\nu_{0}} \sim 0.070 
\left(\frac{1+z}{10}\right)^{2} 
\left(\frac{\Delta \nu}{1 {\rm MHz}}\right),
\eeqa
for the band width $\Delta \nu \sim 1$ MHz.
Note also that the LOFAR
will have $A_{\rm eff}/T_{\rm sys} \sim 
4 \times 10^{6}$ cm$^{2}$ K$^{-1}$ at $\nu \sim 100$ MHz,
so that the required flux is $\sim 10$ times larger
than that for the SKA.

\begin{figure}
\plotone{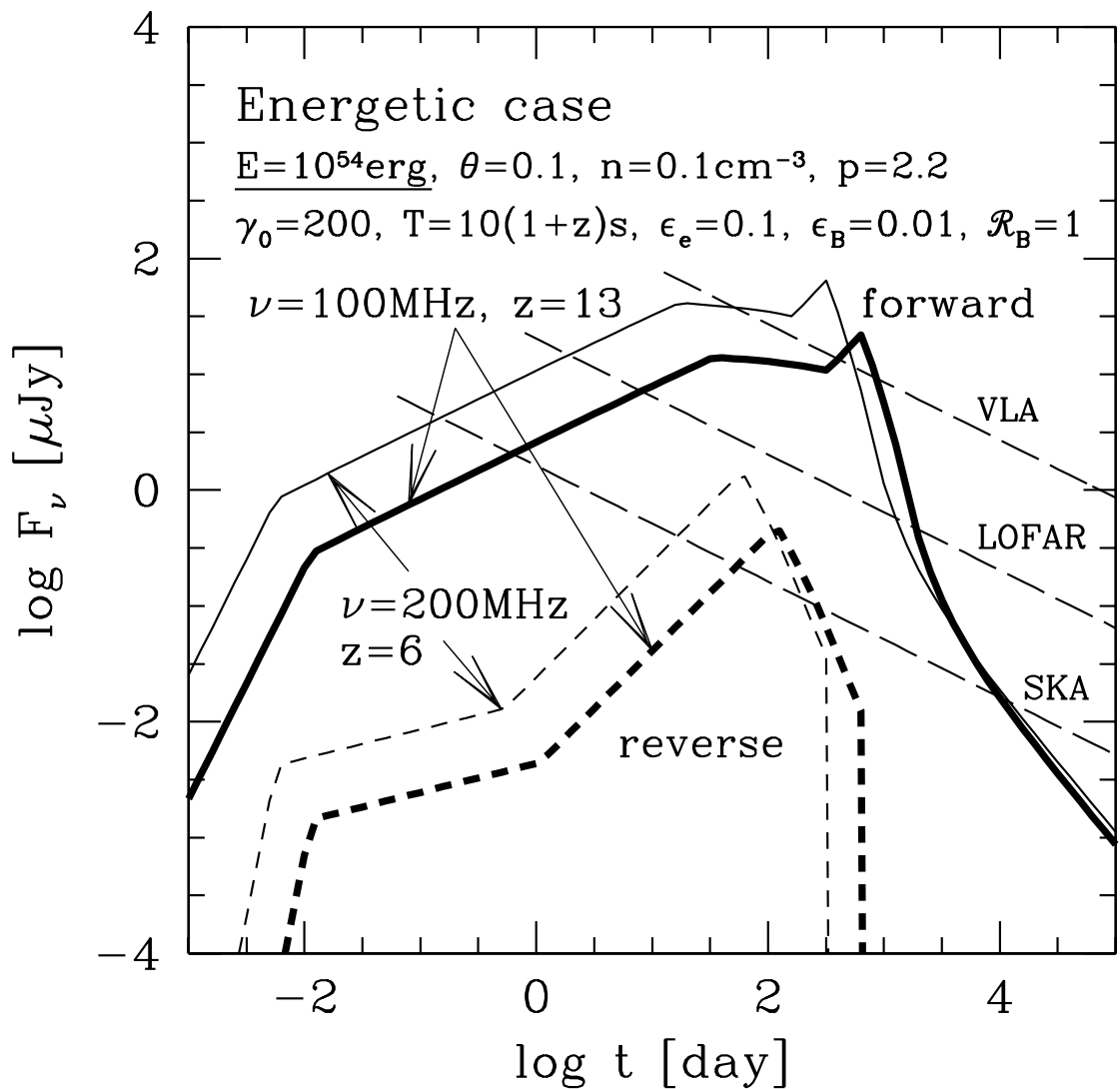}
\caption{\label{fig:ene}
GRB afterglow fluxes at two redshifted 21 cm frequencies, as in 
Figure~\ref{fig:standard} except for the isotropic equivalent energy,
which is taken here as $E=10^{54}$ erg.
}
\end{figure}

Comparing the minimum flux $F_{\nu}^{\rm min}$ of equation 
(\ref{eq:fmin}) with the peak fluxes of the afterglows in 
Figures~\ref{fig:standard}, \ref{fig:ene}, \ref{fig:high},
\ref{fig:long}, \ref{fig:magnetized} and \ref{fig:HN},
we see that in almost all cases there is little or no chance 
to detect the 21 cm line due to the diffuse IGM.
The only one exception is the hypothetical energetic HN case
with isotropic equivalent energy $E\sim 10^{54}$ erg in a
low ambient density $n \siml 1$ cm$^{-3}$,
if the HN occurs at an epoch with low spin temperature 
$T_{S} \siml 0.5 T_{\rm CMB}$.
Although such a HN is not implausible,
there is so far no evidence for it.
We conclude that it is likely to be difficult to detect
21 cm absorption due to the diffuse IGM in the afterglows
discussed here with the currently existing or envisaged 
telescopes. Such detections may require a telescope with 
$\sim 10-100$ times better sensitivity than the SKA.

The 21 cm absorption due to collapsed gas clouds
may have better detection prospects, since the optical depth 
can be high, $\tau \simg 1$, because of the high density contrast.
The number of clouds with optical depth larger than unity
$\tau \simg 1$ may be $0.01$-$0.1$ per redshift, 
while some of host galaxies may have 
$\tau \simg 1$ \citep{furlanetto02}.
In order to resolve the clouds, the bandwidth should be 
narrower than the line width, which is about
\beqa
\Delta \nu \sim 3.9 \left(\frac{T_{S}}{10^{3} {\rm K}}\right)^{1/2}
\left(\frac{1+z}{10}\right)^{-1} {\rm kHz},
\label{eq:dnu}
\eeqa
if the line width is determined by the spin temperature.
We note that a spin temperature of order $T_{S} \sim 10^{3}$ K is 
measured in several damped Ly$\alpha$ systems at low redshift, 
while it may be higher at high redshift \citep{kanekar03}.
If we adopt $\Delta \nu \sim 10$ kHz and $\tau \simg 1$, the required 
minimum source flux is about $\sim 0.02$ mJy from equation (\ref{eq:fmin}).
Then the 21 cm absorption line due to gas clouds may be detected 
in the energetic HN case, as we can see from Figure~\ref{fig:HN},
while from Figures~\ref{fig:standard}, \ref{fig:ene}, \ref{fig:high},
\ref{fig:long} and \ref{fig:magnetized}, it could be marginally 
detected in the other afterglow cases if the ambient density is 
low $n\siml 1$ cm$^{-3}$.

\section{21 cm absorption at $z \siml 6$}\label{sec:z<6}

At redshifts $z \siml 6$, 21 cm absorption is expected predominantly
from neutral gas clouds observed as damped Ly$\alpha$ systems,
rather than from the diffuse IGM, since the latter is almost 
completely ionized. The optical depth of damped Ly$\alpha$ systems 
with the comoving $\rm H_{I}$ column density $N_{\rm H_{I}}$ 
and the spin temperature $T_{S}$ is about
\beqa
\tau=\frac{3 c^{2} h_{\rm P} A_{10} N_{\rm H_{I}}}
{32 \pi k \nu_{0} T_{S} \Delta \nu_{c}}
\sim 6.7 \left(\frac{N_{\rm H_{I}}}{10^{23} {\rm cm}^{-2}}\right)
\left(\frac{T_{S}}{10^{3} {\rm K}}\right)^{-3/2},
\eeqa
where we assumed a comoving line width 
$\Delta \nu_{c}=(1+z) \Delta \nu=39 (T_{S}/10^{3}{\rm K})^{1/2}$ kHz
as in equation (\ref{eq:dnu}) \citep{field59a}.
Thus the optical depth is larger than unity, $\tau \simg 1$,
for an $\rm H_{I}$ column density $N_{\rm H_{I}} \simg 3 \times 10^{22}
(T_{S}/10^{3} {\rm K})^{3/2}$ cm$^{-2}$. The spin temperature of order 
$T_{S} \sim 10^{3}$ K is motivated again on the damped Ly$\alpha$ system
data and discussion of \citet{kanekar03}.

As in the previous section, if we adopt $\Delta \nu \sim 10$ kHz and 
$\tau \simg 1$, the required source flux is about $\sim 0.02$ mJy from 
equation (\ref{eq:fmin}). Thus, the 21 cm line due to the damped 
Ly$\alpha$ systems may be detected in the energetic HN case, as we see 
from Figure~\ref{fig:HN}, while it could be marginally detected in the 
other afterglow cases if the ambient density is low $n\siml 1$ cm$^{-3}$,
from Figures~\ref{fig:standard}, \ref{fig:ene}, \ref{fig:high},
\ref{fig:long} and \ref{fig:magnetized}.
Note that the detection of the 21 cm line at $z \siml 6$
is less difficult than at $z \simg 6$, since the observed 21 cm 
frequency is higher at lower redshifts, and the self-absorbed spectrum 
of the afterglows has a steep rising form, $F_{\nu} \propto \nu^{2}$
or $F_{\nu} \propto \nu^{5/2}$, in equation (\ref{eq:self}).

The number of absorption lines per unit redshift and unit comoving 
$\rm H_{I}$ density is estimated from 
observations of the Ly$\alpha$ systems as
\beqa
\frac{\partial^{2} N}{\partial z \partial N_{\rm H_{I}}}\sim 
C_{\rm H_{I}} N_{\rm H_{I}}^{-\lambda},
\eeqa
where $C_{\rm H_{I}}\sim 10^{5.2}$ 
and $\lambda \sim 1.33$ \citep[e.g.,][]{Penton:2004},
and the redshift evolution does not seem to be strong 
\citep[e.g.,][]{storrie00}.
Therefore the number of the 21 cm line with 
optical depth larger than unity $\tau \simg 1$ 
is about $\sim 0.006 (T_{S}/10^{3} {\rm K})^{3(1-\lambda)/2}$ per redshift.
Since an integration time of order $t_{\rm int} \sim 10$ days
is necessary even for $\tau \sim 1$ clouds, we conclude that
it may be difficult to detect the 21 cm line due to 
the damped Ly$\alpha$ clouds in the $z\siml 6$ range.

\section{Conclusions}\label{sec:con}

We have investigated the radio afterglows of GRBs and HNe
at high redshifts, analyzing their detectability as well as 
their possible usefulness for 21 cm absorption measurements. We 
explored several possible sets of model parameters which are 
plausible at high redshifts (see Table~\ref{tab:model}). 
Our source models include the effects 
associated with reverse as well as forward shock emission, 
jet breaks and sideways expansion of the jet, and the transition 
from a relativistic to non-relativistic expansion regime.

Our results indicate that standard GRB radio afterglows may be
detected with the current VLA up to $z \sim 30$, and energetic
HN afterglows may be detected up to $z \sim 20$,
at high radio frequencies $\sim 5$ GHz.  The proposed SKA should 
be easily able to detect the GRB and HN afterglows 
beyond $z \sim 30$ at frequencies $\sim 5$ GHz.
Other possible effects which may be expected at high redshift, 
such as a larger isotropic equivalent energy, higher external 
density, longer duration GRB and the possibility of magnetized 
ejecta, can all contribute to enhance the detectability of 
afterglows at $\sim 5$ GHz.

We find that it is difficult to detect the 21 cm 
absorption line due to a diffuse neutral IGM, even with the SKA.
A possible exception may be in the case of a very high isotropic
equivalent HN energy $E\sim 10^{54}$ erg, if it occurs in a
low ambient density $n \siml 1$ cm$^{-3}$ at an epoch with low 
spin temperature $T_{S} \siml 0.5 T_{\rm CMB}$.  This case is
very hypothetical, but could be plausible under some conditions.

The 21 cm absorption from collapsed gas clouds along the line of sight,
e.g., from minihalos and galactic disks at $z \simg 6$ and damped 
Ly$\alpha$ systems at $z \siml 6$, is in principle detectable in the 
energetic HN case, and marginally detectable in the GRB afterglow cases 
considered, if the 21 cm optical depth of the clouds is larger than unity 
and the ambient gas density of the GRBs is low, $n \siml 1$ cm$^{-3}$.
However, such high optical depth gas clouds are rare, and their 
detection is not practical given the long integration times needed.
This indicates that if 21 cm absorption is detected in a narrow
range of frequencies, it is likely to originate in the host galaxy 
of the GRB or HN with a high 21 cm optical depth, e.g., in an edge-on 
disk,  star-forming galaxy or protogalaxy, and may thus provide a 
direct measurement of the redshift of the host.

\acknowledgments
We thank M.J.~Rees and S.~Kobayashi for useful discussions.
This work was supported in part by the Eberly Research Funds of Penn State 
and by the Center for Gravitational Wave Physics under grants 
PHY-01-14375 (KI), NASA NAG5-13286, NSF AST 0098416 and the Monell Foundation.

\appendix

\section{Afterglow model}\label{sec:model}

We summarize the afterglow model in this section
\citep[c.f.,][]{zhang03,meszaros02}.
\S~\ref{sec:forward} is for the forward shock, while
\S~\ref{sec:reverse} is for the reverse shock.
A very early evolution of the reverse shock emission is derived
in \S~\ref{sec:early}.
We will neglect the inverse Compton cooling
since almost all phase is in the slow cooling
\citep{sari01}.
We will also neglect the effects of the dispersion delay,
the free-free absorption, and the scintillation \citep[e.g.,][]{ioka03}.

The afterglow evolution is divided into two classes
depending on the shell width $cT(1+z)^{-1}$ \citep{sari95}.
We consider that the shell is thin (thick) if $T<t_{\gamma}$
$(t_{\gamma}<T)$ where
\beqa
t_{\gamma}=\frac{1+z}{4 \gamma_{0}^{2} c}
\left(\frac{3 E}{4\pi \gamma_{0}^{2} n m_{p} c^{2}}\right)^{1/3}
\sim 100 E_{53}^{1/3} \gamma_{0,2}^{-8/3} n^{-1/3} (1+z)\ 
{\rm s}
\label{eq:tgamma}
\eeqa
is the observed time for the forward shock
to collect $\sim \gamma_{0}^{-1}$ of the shell's rest mass,
$E=10^{53} E_{53}$ erg and $\gamma_{0}=10^{2} \gamma_{0,2}$.
The evolution becomes independent of the shell thickness 
in a self-similar manner after the shock crosses the shell at
\beqa
t_{\times}=\max(t_{\gamma},T),
\label{eq:tcross}
\eeqa
where the first (second) value in the bracket
is for the thin (thick) case \citep{kobayashi00b}.
The evolution is also independent of the shell width 
in the very beginning before the time
\beqa
t_{i}=\min(T,t_{N}),
\eeqa
where the first (second) value in the bracket
is for the thin (thick) case,
\beqa
t_{N}=\frac{1+z}{4 \gamma_{0}^{2} c}
\left[\frac{3 E (1+z)}{16 \pi n m_{p} c^{3} T \gamma_{0}^{4}}\right]^{1/2}
\sim 100 E_{53}^{1/2} \gamma_{0,2}^{-4} n^{-1/2} T_{2}^{-1/2}\
{\rm s}
\eeqa
is the time for the reverse shock to be relativistic,
and $T=10^{2} T_{2}$ corresponds to 
the spreading time of the shell \citep{sari95}.
Thus the evolution depends on the shell width 
in the interval $t_{i}<t<t_{\times}$.

\subsection{Forward shock}\label{sec:forward}

First let us consider the forward shock.
We may divide the problem into two parts: dynamics and radiation.

{\it Dynamics}:
In the beginning $t<t_{i}$, the swept mass is too small to affect the system and
the Lorentz factor is constant,
\beqa
\gamma(t) \sim \gamma_{0},
\eeqa
as functions of the observed time $t$.
In the interval $t_{i}<t<t_{\times}$, 
the Lorentz factor is still constant
for thin shells, $\gamma(t) \sim \gamma_{0}$,
while the deceleration begins for thick shells,
\beqa
\gamma(t) \sim \gamma_{0} (t/t_{N})^{-1/4},
\eeqa
since the reverse shock becomes relativistic \citep{sari95,sari97}.
After entering the self-similar phase, $t_{\times}<t$, the hydrodynamics is
determined by the adiabatic condition 
$E \sim 4 \pi \gamma^{2} R^{3} n m_{p} c^{2}/3$ and
the relation $R \sim 4 \gamma^{2} ct/(1+z)$ in equation (\ref{eq:R}) as
\beqa
\gamma(t) \sim [3 E (1+z)^{3}/256 \pi n m_{p} c^{5} t^{3}]^{1/8},
\label{eq:gself}
\label{eq:BM}
\eeqa
irrespective of the shell width.
The Lorentz factor drops below $\theta^{-1}$ 
around the time
\beqa
t_{\theta} \sim \frac{1+z}{4 \theta^{-2} c} R_{\theta}
\sim 5 \times 10^{4} E_{53}^{1/3} \theta_{-1}^{8/3} n^{-1/3} (1+z)\ 
{\rm s},
\label{eq:ttheta}
\eeqa
where 
\beqa
R_{\theta} \sim (3 E \theta^{2}/4 \pi n m_{p} c^{2})^{1/3}.
\eeqa
Then the jet starts to expand sideways \citep{rhoads99,sari99b},
so that the dynamics changes as
\beqa
\gamma(t) 
\sim \theta^{-1} (t/t_{\theta})^{-1/2}.
\label{eq:gth}
\eeqa
Finally the shock velocity becomes non-relativistic 
\citep{waxman98,frail00}
around the time 
\beqa
t_{NR}=\frac{1+z}{4 \gamma_{NR}^{2} c} R_{\theta}
\sim 2 \times 10^{6} E_{52}^{1/3} \theta_{-1}^{2/3} n^{-1/3} (1+z)\
{\rm s},
\eeqa
where $\gamma_{NR}=(1-\beta_{NR}^{2})^{-1/2}$ and we adopt
the non-relativistic velocity $\beta_{NR}=2^{-1/2}$.
The flow becomes nearly spherical, so that
the dynamics is described by the self-similar Sedov solution as
\beqa
\beta(t) \sim \beta_{NR} (t/t_{NR})^{-3/5}.
\label{eq:sedov}
\eeqa
For simplicity we use the shock radius given by
\beqa
R(t) \sim 4 [\gamma(t)]^{2} c t/(1+z),
\label{eq:R}
\eeqa
for the relativistic case \citep{sari98,waxman97,panaitescu98},
while $R(t) \propto \beta(t) t$ for the non-relativistic case.

{\it Radiation}:
Once the dynamics above determine $\gamma(t)$ and $R(t)$,
we may estimate the synchrotron spectrum as follows.
The synchrotron spectrum from relativistic electrons in a power-law distribution
usually has four power-law segments with three breaks
at the self-absorption frequency $\nu_{a,f}$, 
the characteristic synchrotron frequency $\nu_{m,f}$,
and the cooling frequency $\nu_{c,f}$ \citep{sari98b}.
If we temporarily neglect the self-absorption, 
the observed flux is given by
\beqa
F_{\nu,f}=\left\{
\begin{array}{ll}
(\nu/\nu_{m,f})^{1/3} F_{\nu,{\rm max},f},& \nu<\nu_{m,f},\\
(\nu/\nu_{m,f})^{-(p-1)/2} F_{\nu,{\rm max},f},& \nu_{m,f}<\nu<\nu_{c,f},\\
(\nu_{c,f}/\nu_{m,f})^{-(p-1)/2} (\nu/\nu_{c,f})^{-p/2} 
F_{\nu,{\rm max},f},& \nu_{c,f}<\nu,
\end{array}\right.
\label{eq:slow}
\eeqa
in the slow cooling case $\nu_{m,f}<\nu_{c,f}$, and
\beqa
F_{\nu,f}=\left\{
\begin{array}{ll}
(\nu/\nu_{c,f})^{1/3} F_{\nu,{\rm max},f},& \nu<\nu_{c,f},\\
(\nu/\nu_{c,f})^{-1/2} F_{\nu,{\rm max},f},& \nu_{c,f}<\nu<\nu_{m,f},\\
(\nu_{m,f}/\nu_{c,f})^{-1/2} (\nu/\nu_{m,f})^{-p/2} 
F_{\nu,{\rm max},f},& \nu_{m,f}<\nu,
\end{array}\right.
\label{eq:fast}
\eeqa
in the fast cooling case $\nu_{c,f}<\nu_{m,f}$,
where
\beqa
F_{\nu,{\rm max},f}=\frac{\sigma_{T} m_{e} c^{2}}{9 q_{e}}
\frac{R^{3} n B_{f} \gamma (1+z)}{D^{2}}
\eeqa
is the observed peak flux at the luminosity distance
$D=(1+z) \int_{0}^{z} (1+z) |dt/dz| dz$ with
$|dt/dz|^{-1}=(1+z) H_{0} [\Omega_{m}(1+z)^{3}+\Omega_{\Lambda}]^{1/2}$,
and
\beqa
B_{f}=[32 \pi m_{p} \epsilon_{B,f} n \gamma (\gamma-1)]^{1/2} c,
\label{eq:Bf}
\eeqa
is the comoving magnetic field.
The characteristic synchrotron frequency $\nu_{m,f}$
and the cooling frequency $\nu_{c,f}$ are given by
$\nu(\gamma_{m,f})$ and $\nu(\gamma_{c,f})$, respectively,
where
\beqa
\nu(\gamma_{e})=\frac{\gamma \gamma_{e}^{2} q_{e} B}{2\pi m_{e} c (1+z)},
\eeqa
is the frequency at which electrons 
with the Lorentz factor $\gamma_{e}$ radiate, and
\beqa
\gamma_{m,f}=\epsilon_{e,f} \frac{p-2}{p-1} \frac{m_{p}}{m_{e}} (\gamma-1),
\label{eq:gmf}
\quad
\gamma_{c,f}=\frac{6\pi m_{e} c (1+z)}{\sigma_{T} \gamma B_{f}^{2} t}.
\eeqa
Note that we may use $B_{f}$ and $\gamma_{m,f}$
in equations (\ref{eq:Bf}) and (\ref{eq:gmf}) even in the non-relativistic phase.

The synchrotron self-absorption limits the flux below the
blackbody emission with the electron temperature
\citep[e.g.,][]{sari99a}. This is given by
\beqa
F_{\nu,\rm BB}=2\pi \nu^{2} \gamma \gamma_{e} m_{e}
(R_{\perp}/D)^{2} (1+z)^{3},
\label{eq:self}
\eeqa
where $\nu$ is the observed frequency,
$R_{\perp}$ is the observed size of the afterglow,
and $\gamma_{e}$ is the typical Lorentz factor of 
the electrons emitting at $\nu$.
For simplicity we use
$R_{\perp}(t)=4 \gamma(t) c t/(1+z)$ for the relativistic case,
while $R_{\perp}(t) \propto \beta(t) t$ for the non-relativistic case.

\subsection{Reverse shock}\label{sec:reverse}

{\it Dynamics}:
In the initial stage $t<t_{i}$, 
the Lorentz factor is constant $\gamma \sim \gamma_{0}$.
In the interval $t_{i}<t<t_{\times}$,
we have $\gamma \sim \gamma_{0}$ for thin shells,
while $\gamma \sim \gamma_{0} (t/t_{N})^{-1/4}$ for thick shells.
These evolutions are the same as that of the forward shock.
After the shell crossing $t_{\times}<t$, 
the evolution is approximately given by
\beqa
\gamma \sim \gamma_{\times} (t/t_{\times})^{-1/2},
\label{eq:grev}
\eeqa
where $\gamma_{\times}=\min(\gamma_{0},\gamma_{T})$ is the
Lorentz factor at the shell crossing and
\beqa
\gamma_{T} \sim 100 E_{53}^{1/8} n^{-1/8} T_{2}^{-3/8} (1+z)^{3/8},
\eeqa
is given by $\gamma(T)$ from equation (\ref{eq:gself})
\citep{zhang03b,kobayashi00}.
A reliable calculation after $\gamma$ drops below $\theta^{-1}$
at $t_{\theta}<t$ requires full numerical simulations and 
is beyond the scope of this paper.
Here we use equation (\ref{eq:grev}) till 
the velocity becomes non-relativistic $\gamma \sim \gamma_{NR}$,
and use the time dependence of the Sedov solution 
in equation (\ref{eq:sedov}) after that.
Since the evolution in equation (\ref{eq:grev})
is the same as that of the sideways expansion
in equation (\ref{eq:gth}), this treatment might not
be completely wrong.
For simplicity we use the shock radius in equation (\ref{eq:R})
for all cases.

{\it Radiation}:
At the shell crossing $t \sim t_{\times}$,
the forward and reverse shocks have
the same Lorentz factor $\gamma_{\times}$, 
energy density $e$, and total energy $e V$.
But these shocks have
the different relative Lorentz factor across the shock $\bar \gamma$,
i.e., the forward shock has $\bar \gamma_{f} \sim \gamma_{\times}$
and the reverse shock has $\bar \gamma_{r} \sim \gamma_{0}/\gamma_{\times}$.
These shocks may also have 
different magnetization parameters $\epsilon_{B,f} \ne \epsilon_{B,r}$
since the ejected shell may carry magnetic fields
from the central source \citep[e.g.,][]{zhang03b}.
Then we can show
\beqa
\frac{\gamma_{m,r}(t_{\times})}{\gamma_{m,f}(t_{\times})}
\sim \frac{1}{\hat \gamma},
\quad
\frac{\gamma_{c,r}(t_{\times})}{\gamma_{c,f}(t_{\times})}
\sim \frac{1}{{\cal R}_{B}^{2}},
\eeqa
since $\gamma_{m} \propto \bar \gamma$,
$\gamma_{c} \propto \gamma^{-1} B^{-2} t^{-1}$
and $B \propto \epsilon_{B} e^{1/2}$, where we define
\beqa
\hat \gamma = \frac{\gamma_{\times}^{2}}{\gamma_{0}},
\quad
{\cal R}_{B} = \frac{B_{r}}{B_{f}}=
\left(\frac{\epsilon_{B,r}}{\epsilon_{B,f}}\right)^{1/2}.
\label{eq:rB}
\eeqa
We can also show
\beqa
\frac{\nu_{m,r}(t_{\times})}{\nu_{m,f}(t_{\times})}\sim 
\frac{{\cal R}_{B}}{\hat \gamma^{2}},
\quad
\frac{\nu_{c,r}(t_{\times})}{\nu_{c,f}(t_{\times})}\sim 
\frac{1}{{\cal R}_{B}^{3}},
\quad
\frac{F_{\nu,{\rm max},r}(t_{\times})}{F_{\nu,{\rm max},f}(t_{\times})}
\sim \hat \gamma
{\cal R}_{B},
\eeqa
since $\nu_{m} \propto \gamma B \gamma_{m}^{2}$,
$\nu_{c} \propto \gamma^{-1} B^{-3} t^{-2}$,
$F_{\nu,{\rm max}} \propto \gamma B N_{e}$
and $N_{e} \propto e V \bar \gamma^{-1}$,
where $N_{e}$ is the total number of emitting electrons.

As in the forward shock case, the synchrotron spectrum is given by
equations (\ref{eq:slow}), (\ref{eq:fast}) and (\ref{eq:self})
replacing the subscript $f$ with $r$,
although the emission above $\nu_{c,r}$ disappears after
the shell crossing $t_{\times}<t$ (see below).
Thus the light curves can be calculated by specifying the temporal evolutions
of $\nu_{m}$, $\nu_{c}$ and $F_{\nu,{\rm max}}$.
The evolutions of $\gamma_{m}$ and $\gamma_{c}$
are also useful for estimating the self-absorption.
In the initial stage $t<t_{i}$,
the evolutions are given by
\beqa
\nu_{m,r} \propto t^{4},\quad
\nu_{c,r} \propto t^{-2},\quad
F_{\nu,{\rm max},r} \propto t^{2},
\label{eq:nuini}
\\
\gamma_{m,r} \propto t^{2},\quad
\gamma_{c,r} \propto t^{-1},
\label{eq:gini}
\eeqa
as shown in \S~\ref{sec:early}.
In the interval $t_{i}<t<t_{\times}$, thin shells have the dependence
\beqa
\nu_{m,r} \propto t^{6},\quad
\nu_{c,r} \propto t^{-2},\quad
F_{\nu,{\rm max},r} \propto t^{3/2},
\\
\gamma_{m,r} \propto t^{3},\quad
\gamma_{c,r} \propto t^{-1},
\eeqa
while the thick shells have
\beqa
\nu_{m,r} \propto t^{0},\quad
\nu_{c,r} \propto t^{-1},\quad
F_{\nu,{\rm max},r} \propto t^{1/2},
\\
\gamma_{m,r} \propto t^{1/4},\quad
\gamma_{c,r} \propto t^{-1/4}.
\eeqa
After the shell crossing $t_{\times}<t$,
no electron is shocked and electrons only cool adiabatically, so that
the evolutions are given by
\beqa
\nu_{m,r} \propto t^{-3/2},\quad
\nu_{c,r} \propto t^{-3/2},\quad
F_{\nu,{\rm max},r} \propto t^{-1},
\\
\gamma_{m,r} \propto t^{-1/4},\quad
\gamma_{c,r} \propto t^{-1/4},
\eeqa
and the emission above $\nu_{c,r}$ disappears
\citep{zhang03b,kobayashi00,gou04}.
In the Sedov phase we use
\beqa
\nu_{m,r} \propto t^{-3},\quad
\nu_{c,r} \propto t^{-3},\quad
F_{\nu,{\rm max},r} \propto t^{-3/5},
\\
\gamma_{m,r} \propto t^{-6/5},\quad
\gamma_{c,r} \propto t^{-6/5},
\eeqa
by noting that the energy density drops as 
$\propto \beta^{2} \propto t^{-5/6}$.

\subsection{Very early reverse shock}\label{sec:early}

Let us derive the evolution of the reverse shock emission 
in the initial stage $t<t_{i}$.
The evolution is independent of the shell width.
The Lorentz factor is constant $\gamma \sim \gamma_{0}$
since the reverse shock is non-relativistic $\bar \gamma_{r} -1 \ll 1$
\citep{sari95}.
This together with the constant ambient density
leads to a constant energy density in the forward and reverse shock
$e_{r} \sim {\rm const}$.
Since the shell spreading has not started,
the shell density decreases as $n_{r} \propto R^{-2} \propto t^{-2}$.
Then the relative velocity across the reverse shock is 
$\beta_{r} \propto (\bar \gamma_{r}-1)^{1/2} 
\propto e_{r}^{1/2} n_{r}^{-1/2} \propto t$,
and hence the total number of emitting electrons increases as 
$N_{e} \propto \beta_{r} t \propto t^{2}$.
Recalling $\gamma_{m} \propto e_{r}/n_{r}$, 
$\gamma_{c} \propto \gamma^{-1} B^{-2} t^{-1}$
and $B \propto e_{r}^{1/2}$,
we have the time dependence in equation (\ref{eq:gini}).
Similarly we can derive equation (\ref{eq:nuini})
since $\nu_{m} \propto \gamma B \gamma_{m}^{2}$,
$\nu_{c} \propto \gamma^{-1} B^{-3} t^{-2}$ and
$F_{\nu,{\rm max}} \propto \gamma B N_{e}$.

%
%

\newpage
\clearpage
\begin{deluxetable}{cccccccccccc}
\tabletypesize{\footnotesize}
\tablecaption{\label{tab:model}
Model parameters and some results:
the isotropic equivalent energy $E$ erg,
the opening half-angle $\theta$,
the ISM density $n$ cm$^{-3}$,
the spectral index $p$,
the initial Lorentz factor $\gamma_{0}$,
the duration in the source frame $T/(1+z)$ s,
the energy fraction of electrons $\epsilon_{e}$,
the energy fraction of magnetic fields $\epsilon_{B}$,
the ratio of reverse to forward magnetic field ${\cal R}_{B}$,
the maximum redshift out to which these afterglows
could be detected with the VLA $z_{\rm VLA}$,
and the peak flux at the redshifted 21 cm frequency,
$\nu=1420/(1+z)$ MHz, for $z \simg 6$, 
$F_{\nu=\frac{1420}{1+z}{\rm MHz}}^{\rm peak}$ $\mu$Jy.
}
\tablecolumns{6}
\tablewidth{0pc}
\tablehead{
\colhead{model} & 
\colhead{$E$ [erg]} & 
\colhead{$\theta$} &
\colhead{$n$ [cm$^{-3}$]} & 
\colhead{$p$} & 
\colhead{$\gamma_{0}$} & 
\colhead{$\frac{T}{1+z}$ [s]} & 
\colhead{$\epsilon_{e}$} & 
\colhead{$\epsilon_{B}$} & 
\colhead{${\cal R}_{B}$} & 
\colhead{$z_{\rm VLA}$} &
\colhead{$F_{\nu=\frac{1420}{1+z}{\rm MHz}}^{\rm peak}$ [$\mu$Jy]}}
\startdata
Standard GRB & $10^{53}$ & 0.1 & 0.1 & 2.2 & 200 & 10 & 0.1 & 0.01 & 1 &
 $\sim 30$ & 1-10 \\
Energetic GRB & $10^{54}$ & 0.1 & 0.1 & 2.2 & 200 & 10 & 0.1 & 0.01 & 1 &
$>30$ & 10-$10^{2}$ \\
High density GRB & $10^{53}$ & 0.1 & $10^{2}$ & 2.2 & 200 & 10 & 0.1 & 0.01 & 1 &
$>30$ & $10^{-1}$-1 \\
Long duration GRB & $10^{53}$ & 0.1 & 0.1 & 2.2 & 200 & $10^{3}$ & 0.1 & 0.01 & 1 &
$>30$ & 1-10 \\
Magnetized GRB & $10^{53}$ & 0.1 & 0.1 & 2.2 & 200 & 10 & 0.1 & 0.01 & 5 &
$\sim 30$ & 1-10 \\
Hypernova & $10^{54}$ & $1/\sqrt{2}$ & 0.1 & 2.2 & 2 & 10 & 0.1 & 0.01 & 1 &
$\sim 20$ & $10^{2}$-$10^{3}$ \\
\enddata
\end{deluxetable}

%
%

\newpage

\newpage

\newpage

\newpage

\newpage

\newpage

\newpage

\newpage

\newpage

\newpage

\newpage

\newpage

\end{document}